\documentclass[preprintnumbers, floatfix, onecolumn, preprintnumbers, letterpaper, superscriptaddress,nofootinbib]{revtex4}
\usepackage{amsfonts}
\usepackage{amsmath}
\usepackage{amssymb,epsf}
\usepackage{latexsym}
\usepackage{graphicx,epsfig}
\usepackage{amssymb}
\usepackage{subfigure}
\usepackage[colorlinks=true,citecolor=blue,linkcolor=blue,urlcolor=black]{hyperref}
\graphicspath{{Images/}}

\begin{document}

\title{Thermodynamics, phase transitions and Ruppeiner geometry\\
for Einstein-dilaton Lifshitz black holes\\
in the presence of Maxwell and Born-Infeld electrodynamics}
\author{M. Kord Zangeneh}
\email{mkzangeneh@shirazu.ac.ir}
\affiliation{Center of Astronomy and Astrophysics, Department of Physics and Astronomy,
Shanghai Jiao Tong University, Shanghai 200240, China}
\affiliation{Physics Department and Biruni Observatory, Shiraz University, Shiraz 71454,
Iran}
\author{A. Dehyadegari}
\affiliation{Physics Department and Biruni Observatory, Shiraz University, Shiraz 71454,
Iran}
\author{M. R. Mehdizadeh}
\email{mehdizadeh.mr@uk.ac.ir}
\affiliation{Department of Physics, Shahid Bahonar University, P.O. Box 76175, Kerman,
Iran}
\affiliation{Research Institute for Astrophysics and Astronomy of Maragha (RIAAM), P.O.
Box 55134-441, Maragha, Iran}
\author{B. Wang}
\email{wang\_b@sjtu.edu.cn}
\affiliation{Center of Astronomy and Astrophysics, Department of Physics and Astronomy,
Shanghai Jiao Tong University, Shanghai 200240, China}
\author{A. Sheykhi}
\email{asheykhi@shirazu.ac.ir}
\affiliation{Physics Department and Biruni Observatory, Shiraz University, Shiraz 71454,
Iran}
\affiliation{Research Institute for Astrophysics and Astronomy of Maragha (RIAAM), P.O.
Box 55134-441, Maragha, Iran}

\begin{abstract}
In this paper, we first obtain the higher dimensional dilaton-Lifshitz black
hole solutions in the presence of Born-Infeld (BI) electrodynamics. We find
that there are two different solutions for $z=n+1$ and $z\neq n+1$ cases
where $z$ is dynamical critical exponent and $n$ is the number of spatial
dimensions. Calculating the conserved and thermodynamical quantities, we
show that the first law of thermodynamics is satisfied for both cases. Then,
we turn to study different phase transitions for our Lifshitz black holes.
We start with Hawking-Page phase transition and explore the effects of
different parameters of our model on it for both linearly and BI charged
cases. After that, we discuss the phase transitions inside the black holes.
We present the improved Davies quantities and prove that the phase
transition points shown by them are in coincident with Ruppeiner ones. We
show that the zero temperature phase transitions are transitions on radiance
properties of black holes by using Landau-Lifshitz theory of thermodynamic
fluctuations. Next, we turn to study Ruppeiner geometry (thermodynamic
geometry) for our solutions. We investigate thermal stability, interaction
type of possible black hole molecules and phase transitions of our solutions
for linearly and BI charged cases separately. For linearly charged case, we
show that there are no phase transition at finite temperature for the case $%
z\geq 2$. For $z<2$, it is found that the number of finite temperature phase
transition points depends on the value of black hole charge and is not more
than two. When we have two finite temperature phase transition points, there
are no thermally stable black hole between these two points and we have
discontinues small/large black hole phase transitions. As expected, for
small black holes, we observe finite magnitude for Ruppeiner invariant which
shows the finite correlation between possible black hole molecules while for
large black holes, the correlation is very small. Finally, we study the
Ruppeiner geometry and thermal stability of BI charged Lifshtiz black holes
for different values of $z$. We observe that small black holes are thermally
unstable in some situations. Also, the behavior of correlation between
possible black hole molecules for large black holes is the same as linearly
charged case. In both linearly and BI charged cases, for some choices of
parameters, the black hole systems behave like a Van der Waals gas near
transition point.
\end{abstract}

\maketitle


\section{Introduction}

It has been over forty years since Bekenstein and Hawking first disclosed
that black hole can be considered as a thermodynamic system, with
characteristic temperature and entropy \cite{Th1,Th2,Th3,Th4}. Taking into
account the fact that black holes have no hair, there are no classical
degrees of freedom to account for such thermodynamic properties. It is a
general belief that thermodynamic properties of a system may reflect the
statistical mechanics of underlying relevant microscopic degrees of freedom.
But the detailed nature of these microscopic gravitational states has
remained as a mystery. The Bekenstein-Hawking entropy, $S=A/(4\hbar G)$,
depends on both Planck's constant as well as Newtonian gravitational
constant, implying that thermodynamics of black holes may relate quantum
mechanics and gravity. Recently, there have been some progresses on
understanding the microscopic degrees of freedom of the black hole entropy,
for example in string theory \cite{str1,str2,str3} as well as loop quantum
gravity \cite{loop1,loop2,loop3}. But the accounts of the black hole entropy
are not complete and they only work within some particular models and some
special domains where string theory and loop quantum gravity can apply.
Besides, despite counting very different states, many inequivalent
approaches to quantum gravity obtain identical results and it is not clear
why any counting of microstates should reproduce the same Bekenstein-Hawking
entropy \cite{Carlip}. The statistical mechanical description of black hole
entropy is still not elegant.

On the other side, black hole can be heated or cooled though absorption and
evaporation processes. According to Boltzmann's insight, if a system can be
heated, it must have microscopic structures. Recently, in \cite{micstr},
possible microscopic structures of a charged anti-de Sitter black hole have
been studied and some kind of interactions between possible micromolecules
have been investigated by an interesting physical tool, the Ruppeiner
geometry. Derived from the thermodynamic fluctuation theory, the Ruppeiner
geometry \cite{Rup0,Rup1} is considered powerful to explore the possible
interactions between black hole microscopic structures. The sign of the
Ruppeiner invariant $\mathfrak{R}$ (the Ricci scalar of Ruppeiner geometry)
was argued to be useful for identifying the physical systems similar to the
Fermi (Bose) ideal gas when $\mathfrak{R}>0$ ($\mathfrak{R}<0$) or the
classical ideal gas when $\mathfrak{R}=0$ \cite{Osh}. Besides, the sign of
the Ruppeiner invariant $\mathfrak{R}$ can further be used to interpret the
type of the dominated interaction between molecules of a thermodynamic
system. When $\mathfrak{R}>0$, there is a repulsive interaction between
molecules, when $\mathfrak{R}<0$ the interaction is attractive, and for $%
\mathfrak{R}=0$ there is no interaction in the microstates \cite%
{Rup2,Rup3,Rup4}. Moreover, the magnitude of the Ruppeiner invariant $%
\left\vert \mathfrak{R}\right\vert $ measures the average number of
correlated Planck areas on the event horizon for a black hole system \cite%
{Rup05}. For a review on the description of the Ruppeiner geometry in black
hole systems, we refer to \cite{Rup5,Rup6} and references therein. Further
studies on molecular interactions of black holes, based on the Ruppeiner
geometry, have been carried out in \cite{micstr,comment,shdehmon}.

Phase transition is another interesting topic in black holes thermodynamics.
Davies discussed thermodynamic phase transition of the black holes by
looking at the behavior of the heat capacity \cite{Davies1,Davies2,Davies3}.
He claimed that the discontinuity of the heat capacity marks the second
order phase transition in black holes. However, it was argued that physical
properties do not show any speciality at this discontinuity point if
compared with other heat capacity values, for example the regularity of the
event horizon is not lost and the black hole internal state remains
uninfluenced \cite{soko}. Thus, it is hard to accept the discontinuity point
of the heat capacity as a true physical point of the phase transition.
Employing the Landau-Lifshitz theory of thermodynamic fluctuations \cite%
{LL1,LL2}, Pavon and Rubi gave a deep understanding of the black hole phase
transition \cite{pavon1,pavon2}. They found that some second moments in the
fluctuation of relevant thermodynamic quantities diverge when the black hole
becomes extreme. This divergence shows that the thermodynamic fluctuation is
tremendous and the rigorous meaning of thermodynamical quantities is broken
down. This is exactly the characteristic of the thermodynamic phase
transition point. At this phase transition point, the Hawking temperature is
zero which indicates that for the extreme black hole there is only
super-radiation but no Hawking radiation, which is in sharp difference from
that of the non-extreme black holes. Black holes phase transition in the
context of Landau-Lifshitz theory have been investigated in \cite{Cai, Wang1}%
. Recently, further differences in dynamical properties before and after the
black hole thermodynamical phase transition has been disclosed in \cite%
{Wang2,Wang3,Wang4}. A question now arises: how we can further understand
this macroscopic thermodynamic phase transition in black hole physics? for
example whether there is a microscopic explanation of this thermodynamic
phase transition. The Ruppeiner geometry is a possible tool we can use to
investigate the thermodynamic phase transitions from microscopic point of
view. This method is safer to determine true phase transitions than other
methods since regardless of microscopic model, $\mathfrak{R}$ has a unique
status in identifying microscopic order (which is at foundation of phase
transitions at microscopic level) from thermodynamics \cite{Rup5,Rup6}. Some
attempts, in this direction, have been reported in \cite%
{RupGeo1,RupGeo11,RupGeo2,RupGeo3,RupGeo4,RupGeo5,RupGeo6,RupGeo61,RupGeo7,RupGeo8,RupGeo9,RupGeo10}%
. In a recent work \cite{RupGeo3}, it was found that the divergence of the
Ruppeiner invariant coincides with the critical point in the phase
transition in a holographic superconductor model. It is interesting to
investigate whether the Ruppeiner geometry \cite{Rup5,Rup6} can present us
further reason to determine which of the thermodynamical discussions
mentioned above is valid for describing the thermodynamical phase
transition. In particular, we would like to explore whether the Davies phase
transition conjecture can reflect some special properties in microstructures
and be in consistent with the Ruppeiner geometry description. If the Davies
conjecture does not have the microscopic explanation, we will further think
about how to improve the Davies conjecture to describe the black hole phase
transition.

We will employ the black hole in Lifshitz spacetime as a configuration to
study our physical problems mentioned above. This spacetime was first
introduced in \cite{Lif}, which respects the anisotropic conformal
transformation $t\rightarrow \lambda ^{z}t$, $\vec{\mathbf{x}}\rightarrow
\lambda \vec{\mathbf{x}}$, where $z$ is dynamical critical exponent. For the
Lifshitz spacetime, it is necessary to include some matter sources such as
massive gauge fields \cite{massgf1,massgf2,massgf3,massgf4,massgf5} or
higher-curvature corrections \cite{hcc} to guarantee the asymptotic behavior
of the Lifshitz black hole. It is difficult to find an analytic Lifshitz
black hole solution for arbitrary $z$, although some attempts have been
performed \cite{alva}. This makes the discussion of thermodynamics for such
a black hole difficult. Fortunately, in Einstein-dilaton gravity with a
massless gauge field, it is possible to find an exact Lifshitz black
solution for arbitrary $z$ \cite{taylor,tario}. This model is suggested in
the low energy limit of string theory \cite{Polch}. While thermodynamical
behaviors of uncharged and charged Einstein-dilaton-Lifshitz black holes
have been revealed in \cite{taylor,peet} and \cite{tario}, respectively,
thermodynamics of uncharged Gauss-Bonnet-dilaton-Lifshitz solution has been
studied in \cite{Kord1}. It is also interesting to study Lifshitz black hole
solutions in the presence of other gauge fields such as the power-law
Maxwell field \cite{PL}, the logarithmic \cite{Log} and exponential \cite%
{Exp} nonlinear electrodynamics. For example, thermodynamics and thermal and
dynamical stabilities of Einstein-dilaton-Lifshitz solutions in the presence
of power-law Maxwell field have been studied in \cite{Kord2}. In the context
of AdS/CFT \cite{MW1,MW2,MW3} application, the electrical conductivity were
explored for exponentially \cite{Kord3} and logarithmic \cite{Kord4} charged
Lifshitz solutions. In the present work, we shall consider the Born-Infeld
(BI) nonlinear electrodynamics in the context of Einstein-dilaton-Lifshitz
black holes. The motivation for considering BI-dilaton action comes from the
fact that dynamics of D-branes and some soliton solutions of supergravity is
governed by the Born-Infeld (BI) action \cite{BI1,BI2,BI3,BI4,BI5,BI6}.
Besides, the low energy limit of open superstring theory suggest the BI
electrodynamic action coupled to dilaton field \cite{BI1,BI2,BI3}. It is
surprising to mention that, many years before the appearance of BI action in
superstring theory, in 1930's, this nonlinear electrodynamics was introduced
for the first time, with the aim of solving the infinite self-energy problem
of a point-like charged particle by imposing a maximum strength for the
electromagnetic field \cite{BI7}.

In this paper, we will first look for a general ($n+1$)-dimensional Lifshitz
black hole solution in the context of Einstein-dilaton gravity in the
presence of BI electrodynamics. We will show that the general metric
function has different solutions for $z=n+1$ and $z\neq n+1$ cases. It is
important to note that the difference in the metric function has not been
observed in the previous studies on Lifshitz-dilaton black holes \cite%
{Kord2,Kord3,Kord4}. Based on this general solution, we will study
thermodynamics of Lifshitz-dilaton black holes coupled to linear Maxwell
field and BI nonlinear electrodynamics. We will disclose that the
Hawking-Page phase transition \cite{HP} exists both in the presence of
linear and nonlinear electrodynamics. There are some attempts on study phase
transitions of uncharged Lifshitz solutions for fixed $z$ \cite{HPL1} or in
three \cite{HPL2} and four \cite{HPL3} dimensions. The disclosed
Hawking-Page phase transition in this paper is interesting, since it depends
on different values of $z$ in different spacetime dimensions in the presence
of linear Maxwell and nonlinear BI electrodynamic fields. We will further
concentrate our attention to understand the thermodynamic phase transition
from microstructures. We shall examine the relation between the Ruppeiner
geometry and thermodynamical descriptions of the phase transition such as
the Davies conjecture and the Landau-Lifshitz method. We try to give more
microscopic understanding of the thermodynamical phase transitions in the
black hole system. We explore the thermodynamic geometry (Ruppeiner
geometry) for linearly and nonlinearly charged Lifshitz solutions separately
and disclose the properties of interactions between possible black hole
molecules. Up to our best knowledge, there is no study of thermodynamic
geometry on Lifshitz solutions in literature. Interestingly enough, by
studying Ruppeiner geometry, we have found that our solutions show the Van
der Waals like behavior near critical point in some cases.

The layout of the paper is as follows. In the next section, we provide the
basic field equations and obtain the BI charged Lifshitz-dilaton black hole
solutions. In section \ref{thermo}, we first explore the satisfaction of the
thermodynamics first law for Lifshitz-dilaton black holes in the presence of
BI electrodynamics. Then, we study different phase transitions including
Hawking-Page phase transition and phase transition at zero temperature for
linearly and BI charged cases. In section \ref{RG}, we investigate
thermodynamic geometry of the obtained solutions for linearly and
nonlinearly BI charged cases by adopting the Ruppeiner approach. We finish
with summary and closing remarks in section \ref{Sum}.

\section{Action and asymptotic Lifshitz solutions \label{Sol}}

\begin{figure*}[t]
\centering{%
\subfigure[~$z=1.5$, $m=1.2$, $n=4$]{
   \label{fig1a}\includegraphics[width=.46\textwidth]{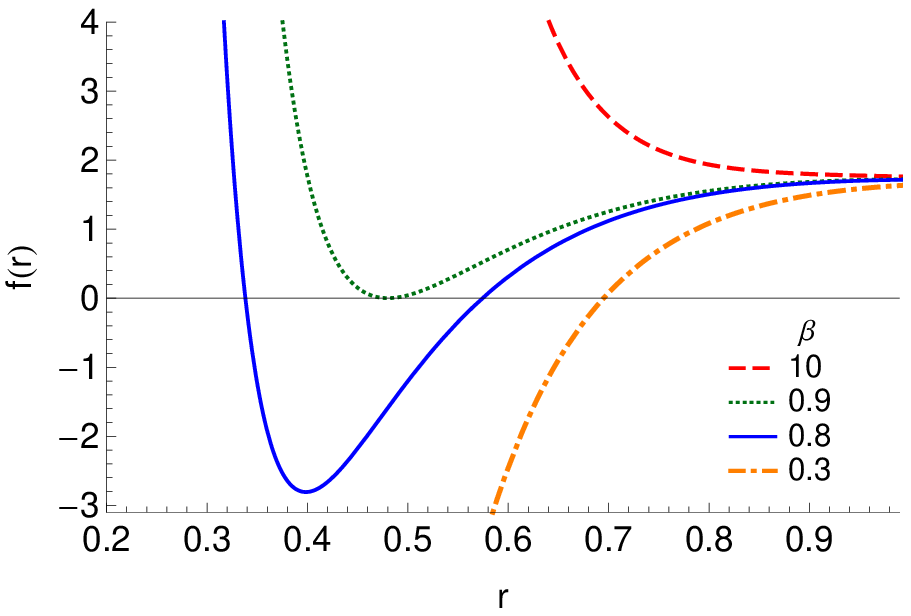}\qquad}} 
\subfigure[~$z=4$, $m=0.45$, $n=3$]{
   \label{fig1b}\includegraphics[width=.46\textwidth]{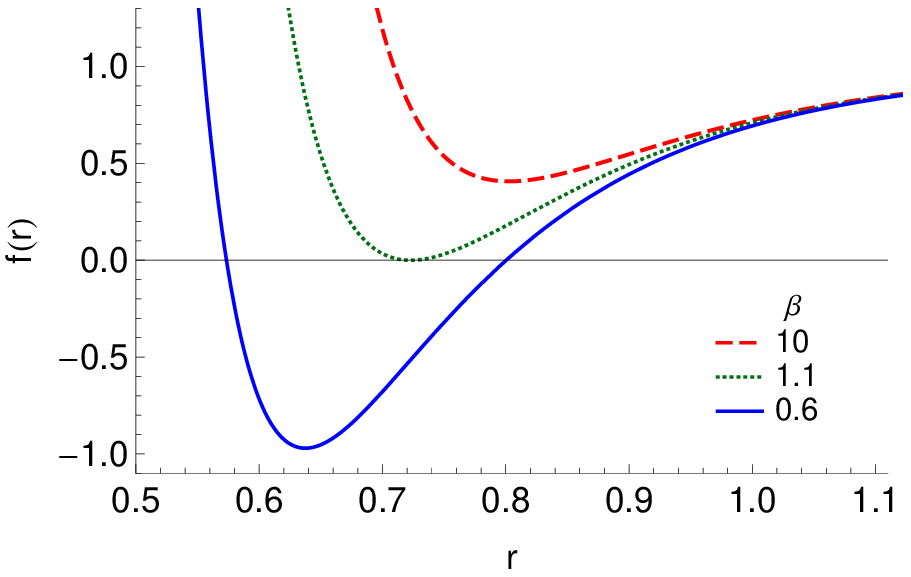}}
\caption{The behavior of $f(r)$ versus $r$ for $l=1$, $b=0.8$ and $q=1.3$.}
\label{fig1}
\end{figure*}

In this section, we intend to obtain exact $(n+1)$-dimensional
dilaton-Lifshitz black holes in the presence of BI nonlinear
electrodynamics. Our ansatz for the line elements of the spacetime is \cite%
{Mann,tario}%
\begin{equation}
ds^{2}=-\frac{r^{2z}f(r)}{l^{2z}}dt^{2}+{\frac{l^{2}dr^{2}}{r^{2}f(r)}}%
+r^{2}d\mathbf{\Omega }_{n-1}^{2},  \label{metric}
\end{equation}%
where $z(\geq 1)$ is dynamical critical exponent and%
\begin{equation*}
d\mathbf{\Omega }_{n-1}^{2}=d\theta _{1}^{2}+\sum\limits_{i=2}^{n-1}d\theta
_{i}^{2}\prod\limits_{j=1}^{i-1}\sin ^{2}\left( \theta _{j}\right) ,
\end{equation*}%
is an ($n-1$)-dimensional hypersurface with constant curvature $(n-1)(n-2)$
and volume $\omega _{n-1}$. As $r\rightarrow \infty $, the line elements (%
\ref{metric}) reduce asymptotically to the Lifshitz spacetime, 
\begin{equation}
ds^{2}=-\frac{r^{2z}dt^{2}}{l^{2z}}+{\frac{l^{2}dr^{2}}{r^{2}}}+r^{2}d%
\mathbf{\Omega }_{n-1}^{2}.  \label{lifmet}
\end{equation}%
On the other side, as it is pointed out above, we would like to consider BI
nonlinear electrodynamics. In the absence of dilaton field, BI Lagrangian
density is written as \cite{BI7}%
\begin{equation}
L=4\beta ^{2}\left( 1-\sqrt{1+\frac{F}{2\beta ^{2}}}\right) ,
\end{equation}%
where $\beta $ is the Born-Infeld parameter related to the Regge slope $%
\alpha ^{\prime }$ as $\beta =1/\left( 2\pi \alpha ^{\prime }\right) $. $%
F=F_{\mu \nu }F^{\mu \nu }$ is Maxwell invariant in which $F_{\mu \nu
}=\partial _{\lbrack \mu }A_{\nu ]}$ where $A_{\mu }$ is electromagnetic
potential. One of the effects of presence of dilaton field is its coupling
with electromagnetic field. Thus, in the presence of dilaton field we deal
with a modified form for BI Lagrangian density including its coupling with
dilaton scalar field $\Phi $ \cite{BID1,BID2}%
\begin{equation}
L(F,\Phi )=4\beta ^{2}e^{4\lambda \Phi /(n-1)}\left( 1-\sqrt{1+\frac{%
e^{-8\lambda \Phi /(n-1)}F}{2\beta ^{2}}}\right) ,  \label{BI Lag}
\end{equation}%
where $\lambda $ is a constant. The Lagrangian density of string-generated
Einstein--dilaton model \cite{Polch} with two Maxwell gauge fields \cite%
{tario} in the presence of BI electrodynamics can be written in Einstein
frame as%
\begin{equation}
\mathcal{L}=\frac{1}{16\pi }\left\{ \mathcal{R}-\frac{4}{n-1}(\nabla \Phi
)^{2}-2\Lambda -\sum\limits_{i=1}^{2}e^{-4\Phi \lambda
_{i}/(n-1)}H_{i}+L(F,\Phi )\right\} ,  \label{Lag}
\end{equation}%
where $\mathcal{R}$ is Ricci scalar and $\Lambda $ and $\lambda _{i}$'s are
some constants. In Lagrangian (\ref{Lag}), $H_{i}=\left( H_{i}\right) _{\mu
\nu }\left( H_{i}\right) ^{\mu \nu }$ where $\left( H_{i}\right) _{\mu \nu
}=\partial _{\lbrack \mu }\left( B_{i}\right) _{\nu ]}$ and $\left(
B_{i}\right) _{\mu } $ is gauge potential. In the large $\beta $ limit, $%
\mathcal{L}$ recovers the Einstein-dilaton-Maxwell Lagrangian in its leading
order \cite{tario,Kord2}%
\begin{equation}
\lim_{\beta \rightarrow \infty }16\pi \mathcal{L}=\cdots -e^{-4\lambda \Phi
/(n-1)}F+\frac{e^{-12\lambda \Phi /(n-1)}F^{2}}{8\beta ^{2}}+O\left( \frac{1%
}{\beta ^{4}}\right) .
\end{equation}%
Varying the action $S=\int_{\mathcal{M}}d^{n+1}x\sqrt{-g}\mathcal{L}$ with
respect to the metric $g_{\mu \nu }$, the dilaton field $\Phi $ and
electromagnetic potentials $A_{\mu }$ and $\left( B_{i}\right) _{\mu }$,
leads us to the following field equations%
\begin{gather}
\mathcal{R}_{\mu \nu }-\frac{g_{\mu \nu }}{n-1}\left\{ 2\Lambda
+2L_{F}F-L(F,\Phi )-\sum\limits_{i=1}^{2}e^{-4\Phi \lambda
_{i}/(n-1)}H_{i}\right\}  \notag \\
-\frac{4}{n-1}\partial _{\mu }\Phi \partial _{\nu }\Phi +2L_{F}F_{\mu
\lambda }F_{\nu }^{\text{ \ }\lambda }-2\sum\limits_{i=1}^{2}e^{-4\lambda
_{i}\Phi /(n-1)}\left( H_{i}\right) _{\mu \lambda }\left( H_{i}\right) _{\nu
}^{\text{ \ }\lambda }=0,  \label{FE1}
\end{gather}%
\begin{eqnarray}
&&\nabla ^{2}\Phi +\frac{n-1}{8}L_{\Phi }+\sum\limits_{i=1}^{2}\frac{{%
\lambda }_{i}}{2}e^{-{4{\lambda }_{i}\Phi }/({n-1})}H_{i}=0,  \label{FE2} \\
&&\triangledown _{\mu }\left( L_{F}F^{\mu \nu }\right) =0,  \label{FE3} \\
&&\triangledown _{\mu }\left( e^{-{4\lambda }_{i}{\Phi }/({n-1})}\left(
H_{i}\right) ^{\mu \nu }\right) =0,  \label{FE4}
\end{eqnarray}%
where we use the convention $X_{Y}=\partial X/\partial Y$. Using the metric
ansatz (\ref{metric}), electromagnetic field equations (\ref{FE3}) and (\ref%
{FE4}) can be solved immediately as%
\begin{eqnarray}
F_{rt} &=&\frac{q\beta e^{4\lambda \Phi /(n-1)}r^{z-n}}{\Upsilon },
\label{Frt} \\
\left( H_{i}\right) _{rt} &=&q_{i}r^{z-n}e^{4\lambda _{i}\Phi /(n-1)},
\label{Hrt}
\end{eqnarray}%
where $\Upsilon =\sqrt{1+q^{2}l^{2z-2}/(\beta ^{2}r^{2n-2})}$, and $\Phi (r)$
can be obtained by subtracting ($tt$) and ($rr$) components of Eq. (\ref{FE1}%
) and solving the resulting equation. We find 
\begin{equation}
\Phi (r)=\frac{(n-1)\sqrt{z-1}}{2}\ln \left( \frac{r}{b}\right) .
\label{Phi}
\end{equation}%
Substituting Eqs. (\ref{Frt}), (\ref{Hrt}) and (\ref{Phi}) in field
equations (\ref{FE1}) and (\ref{FE2}), one can solve the equations for $f(r)$
to obtain%
\begin{equation}
f(r)=\left\{ 
\begin{array}{ll}
1-\frac{m}{r^{n+z-1}}+\frac{(n-2)^{2}l^{2}}{(n+z-3)^{2}r^{2}}+\frac{4\beta
^{2}l^{2}{b}^{2z-2}}{{r}^{2z-2}(n-1)(n-z+1)}-\frac{4\beta ^{2}l^{2}b^{2z-2}}{%
(n-1)r^{n+z-1}}\int \Upsilon r^{n-z}{dr,} & \text{for }z\neq n+1, \\ 
&  \\ 
1-\frac{m}{r^{2n}}+\frac{(n-2)^{2}l^{2}}{4(n-1)^{2}r^{2}}-\frac{4\beta
^{2}b^{2n}l^{2}}{(n-1)^{2}r^{2n}}\left[ 1-\Upsilon +\ln \left( \frac{%
1+\Upsilon }{2}\right) \right] , & \text{for }z=n+1,%
\end{array}%
\right.  \label{f4}
\end{equation}%
where we should set 
\begin{gather}
\lambda =-\sqrt{z-1},\text{ \ \ \ \ }\lambda _{1}=\frac{n-1}{\sqrt{z-1}},%
\text{ \ \ \ \ }\lambda _{2}=\frac{n-2}{\sqrt{z-1}},  \notag \\
q_{1}^{2}=\frac{-\Lambda \left( z-1\right) b^{2(n-1)}}{\left( z+n-2\right)
l^{2(z-1)}},\text{ \ \ \ \ }q_{2}^{2}=\frac{(n-1)(n-2)(z-1)b^{2(n-2)}}{%
2(z+n-3)l^{2(z-1)}},  \notag \\
\Lambda =-\frac{(n+z-1)(n+z-2)}{2l^{2}},  \label{constants}
\end{gather}%
so that the field equations are fully satisfied. In the solution (\ref{f4}), 
$m$ is a constant which is related to the total mass of black brane as we
will see in next section. The integral of the last term of $f(r)$ for $z\neq
n+1$ can be done in terms of hypergeometric function. Thus, $f(r)$ can be
written as%
\begin{align}
f(r)& =1-\frac{m}{r^{n+z-1}}+\frac{(n-2)^{2}l^{2}}{(n+z-3)^{2}r^{2}}+\frac{%
4b^{2z-2}l^{2}\beta ^{2}(1-\Upsilon )}{(n-1)(n-z+1)r^{2z-2}}  \notag \\
& +\frac{4q^{2}b^{2z-2}l^{2z}\Upsilon }{(n+z-3)(n-z+1)r^{2\left(
n+z-2\right) }}\mathbf{F}\left( 1,\frac{2n+z-4}{2n-2},\frac{3n+z-5}{2n-2}%
,1-\Upsilon ^{2}\right) .  \label{fr}
\end{align}%
Note that solution (\ref{fr}) obviously satisfies the fact that $%
f(r)\rightarrow 1$ as $r\rightarrow \infty $ (note that $\mathbf{F}\left(
a,b,c,0\right) =1$). The behavior of $f(r)$ for large $\beta $ is%
\begin{equation}
f(r)=\left\{ 
\begin{array}{ll}
1-\frac{m}{r^{n+z-1}}+\frac{(n-2)^{2}l^{2}}{(n+z-3)^{2}r^{2}}+\frac{%
2q^{2}b^{2z-2}l^{2z}}{\left( n-1\right) \left( n+z-3\right) r^{2n+2z-4}}-%
\frac{q^{4}b^{2z-2}l^{4z-2}}{4\left( n-1\right) \left( 3n+z-5\right) \beta
^{2}r^{4n+2z-6}}+O\left( \frac{1}{\beta ^{4}}\right) {,} & \text{for }z\neq
n+1, \\ 
&  \\ 
1-\frac{m}{r^{2n}}+\frac{(n-2)^{2}l^{2}}{4(n-1)^{2}r^{2}}+\frac{%
q^{2}b^{2n}l^{2n+2}}{\left( n-1\right) ^{2}r^{4n-2}}-\frac{%
q^{4}b^{2n}l^{4n+2}}{8\left( n-1\right) ^{2}\beta ^{2}r^{6n-4}}+O\left( 
\frac{1}{\beta ^{4}}\right), & \text{for }z=n+1.%
\end{array}%
\right.  \label{Lbetaf}
\end{equation}%
%
%
%
%
%
which reproduces the result of \cite{Kord2} for every $z$\ in linear Maxwell
case. The behaviors of the metric function for $z=n+1$\ and $z\neq n+1$\
have been depicted in Figs. \ref{fig1a} and \ref{fig1b} respectively. It is
notable to mention that in the case of $z=n+1$, there is no
Schwartzshild-like black hole since in this case $f(r)$\ goes to positive
infinity as $r$\ goes to zero. However, for $z\neq n+1$, we may have
Schwartzshild-like black hole (dash-dotted line in Fig. \ref{fig1a}) in
addition to nonextreme (solid line) and extreme (dotted line) black holes
and naked singularity (dashed line). For nonextreme case, there are two
inner (Cauchy) and outer (event) horizons. In both Figs. \ref{fig1a} and \ref%
{fig1b}, we see that the larger the nonlinearity parameter $\beta $ is, the
smaller the distance between two inner and outer horizons is so that for
large enough $\beta $'s, we have just one horizon (extreme case) or naked
singularities. The Schwartzshild-like case occurs for lower $\beta $'s in
the case of $z\neq n+1$\ as Fig. \ref{fig1a} shows.

As one can see in (\ref{Lbetaf}), the fourth term in expansions for both $%
z=n+1$\ and $z\neq n+1$\ cases reproduce the charge term of \cite{Kord2} in
linear Maxwell case as one expects. The temperature of the black hole
horizon can be obtained via%
\begin{equation}
T=\frac{1}{2\pi }\left. \sqrt{-\frac{1}{2}\nabla _{b}\chi _{a}\nabla
^{b}\chi ^{a}}\right\vert _{r=r_{+}}  \label{T}
\end{equation}%
where $\chi =\partial _{t}$ is the Killing vector and $r_{+}$ is the radius
of event horizon. Using (\ref{T}), one can calculate the Hawking temperature
as%
\begin{eqnarray}
T &=&\left. \frac{r^{z+1}f^{\prime }}{4\pi l^{z+1}}\right\vert _{r=r_{+}} 
\notag \\
&=&\frac{{(n+z-1){r}_{+}^{z}}}{4\pi l^{z+1}}+\frac{(n-2)^{2}l^{1-z}}{4\pi
(n+z-3)r_{+}^{2-z}}+\frac{\beta ^{2}b^{2z-2}r_{+}^{2-z}\left( 1-\Upsilon
_{+}\right) }{\pi (n-1)l^{z-1}},  \label{Temp}
\end{eqnarray}%
where prime denotes the derivative with respect to $r$ and $\Upsilon
_{+}=\Upsilon \left( r=r_{+}\right) $. Temperature has the same formula (\ref%
{Temp}) for both $z=n+1$\ and $z\neq n+1$\ cases. One can check that for
large $\beta $, (\ref{Temp}) reduces to the temperature of
Einstein-Maxwell-dilaton Lifshitz black holes \cite{Kord2}, namely 
\begin{equation}
T=\frac{(n+z-1)r_{+}^{z}}{4\pi l^{z+1}}+\allowbreak \frac{(n-2)^{2}l^{1-z}}{%
4\pi (n+z-3)r_{+}^{2-z}}-\frac{q^{2}l^{z-1}b^{2z-2}}{2\pi (n-1)r_{+}^{2n+z-4}%
}+\frac{q^{4}l^{3z-3}b^{2z-2}}{8\pi (n-1)r_{+}^{4n+z-6}\beta ^{2}}+O\left( 
\frac{1}{\beta ^{4}}\right) .
\end{equation}
The entropy of the black holes can be calculated by using the area law of
the entropy \cite{Th2,Beck1,Beck2} which is applied to almost all kinds of
black holes in Einstein gravity including dilaton black holes \cite%
{hunt1,hunt2,hunt3,hunt4}. Therefore, the entropy of the black brane per
unit volume $\omega _{n-1}$ becomes%
\begin{equation}
S=\frac{r_{+}^{n-1}}{4}.  \label{entropy}
\end{equation}
Having Eqs. (\ref{Frt}), (\ref{Phi}) and (\ref{constants}) at hand, we can
find electromagnetic gauge potential $A_{t}=\int F_{rt}dr$ in terms of
hypergeometric function as%
\begin{equation}
A_{t}\left( r\right) =-\frac{qb^{2z-2}}{(n+z-3)r^{n+z-3}}\mathbf{F}\left( 
\frac{1}{2},\frac{n+z-3}{2n-2},\frac{3n+z-5}{2n-2},1-\Upsilon ^{2}\right) .
\label{at}
\end{equation}%
The large $\beta $ behavior of gauge potential is in agreement with \cite%
{Kord2}

\begin{equation}
A_{t}\left( r\right) =-\frac{qb^{2z-2}}{(n+z-3)r^{n+z-3}}\mathbf{+}\frac{%
q^{3}b^{2z-2}l^{2z-2}}{\left( 3n+z-5\right) r^{3n+z-5}\beta ^{2}}+O\left( 
\frac{1}{\beta ^{4}}\right) .
\end{equation}%
\qquad In next section, we will study thermodynamics of dilaton Lifshitz
black holes in the presence of BI electrodynamics by seeking for
satisfaction of thermodynamics first law through calculation of conserved
and thermodynamic quantities. We also show that our Lifshitz solutions can
exhibit the Hawking-Page phase transition. Then, we discuss the inside phase
transitions of our Lifshitz black holes.

\section{Thermodynamics of Lifshitz black holes\label{thermo}}

\subsection{First law of thermodynamics}

This subsection is devoted to study the thermodynamics first law for
Lifshitz-dilaton black hole solutions in the presence of BI nonlinear
electrodynamics. As the first step, we calculate the fundamental quantity
for thermodynamics discussions namely mass. For this purpose, we apply the
modified subtraction method of Brown and York (BY) \cite{BY1,BY2,modBY}. In
order to use this method, the metric should be written in the form%
\begin{equation}
ds^{2}=-X(R)dt^{2}+\frac{dR^{2}}{Y(R)}+R^{2}d\mathbf{\Omega }_{n-1}^{2}.
\label{Mets}
\end{equation}%
For our case, it is clear that $R=r$ and thus%
\begin{equation}
X(R)=\frac{r(R)^{2z}f(r(R))}{l^{2z}},\text{ \ \ \ \ }Y(R\mathcal{)}=\frac{%
r(R)^{2}f(r(R))}{l^{2}}.
\end{equation}%
The metric of background is chosen to be the Lifshitz metric (\ref{Mets})
i.e.

\begin{equation}
X_{0}(R)=\frac{r(R)^{2z}}{l^{2z}},\text{ \ \ \ \ }Y_{0}(R)=\frac{r(R)^{2}}{%
l^{2}}.
\end{equation}%
The quasilocal conserved mass can be obtained through

\begin{equation}
M=\frac{1}{8\pi }\int_{\mathcal{B}}d^{2}\varphi \sqrt{\sigma }\left\{ \left(
K_{ab}-Kh_{ab}\right) -\left( K_{ab}^{0}-K^{0}h_{ab}^{0}\right) \right\}
n^{a}\xi ^{b},
\end{equation}%
where $\sigma $ is the determinant of the boundary $\mathcal{B}$ metric, $%
K_{ab}^{0}$ is the background extrinsic curvature, $n^{a}$ is the timelike
unit normal vector to the boundary $\mathcal{B}$ and $\xi ^{b}$ is a
timelike Killing vector field on the boundary surface. Performing the above
modified BY formalism, the mass of the space time per unit volume ${\omega
_{n-1}}$ can be calculated as%
\begin{equation}
M=\frac{(n-1)m}{16\pi l^{z+1}},  \label{Mass}
\end{equation}%
where the mass parameter $m$ can be obtained from the fact that $f(r_{+})=0$
as%
\begin{equation}
m{(r_{+})=}\left\{ 
\begin{array}{ll}
\begin{array}{l}
r_{+}^{n+z-1}+\frac{(n-2)^{2}l^{2}r_{+}^{n+z-3}}{(n+z-3)^{2}}+\frac{%
4b^{2z-2}l^{2}\beta ^{2}(1-\Upsilon _{+})}{(n-1)(n-z+1)r_{+}^{z-n-1}} \\ 
+\frac{4q^{2}b^{2z-2}l^{2z}\Upsilon _{+}}{(n+z-3)(n-z+1)r_{+}^{n+z-3}}%
\mathbf{F}\left( 1,\frac{2n+z-4}{2n-2},\frac{3n+z-5}{2n-2},1-\Upsilon
_{+}^{2}\right)%
\end{array}
& \text{for }z\neq n+1 \\ 
&  \\ 
r_{+}^{2n}+\frac{(n-2)^{2}l^{2}r_{+}^{2n-2}}{4(n-1)^{2}}-\frac{4\beta
^{2}b^{2n}l^{2}}{(n-1)^{2}}\left[ 1-\Upsilon _{+}+\ln \left( \frac{%
1+\Upsilon _{+}}{2}\right) \right] & \text{for }z=n+1%
\end{array}%
\right.  \label{mr+}
\end{equation}%
Now, we turn to calculate the electric charge of the solution. Using the
Gauss law, we can calculate the electric charge via 
\begin{equation}
Q=\frac{\,{1}}{4\pi }\int r^{n-1}L_{F}F_{\mu \nu }n^{\mu }u^{\nu }d{\Omega },
\label{chdef}
\end{equation}%
where%
\begin{equation*}
n^{\mu }=\frac{1}{\sqrt{-g_{tt}}}dt=\frac{l^{z}}{r^{z}\sqrt{f(r)}}dt,\text{
\ \ \ \ }u^{\nu }=\frac{1}{\sqrt{g_{rr}}}dr=\frac{r\sqrt{f(r)}}{l}dr,
\end{equation*}%
are respectively the unit spacelike and timelike normals to the hypersurface
of radius $r$. Using (\ref{chdef}), the charge per unit volume $\omega
_{n-1} $ can be computed as%
\begin{equation}
Q=\frac{ql^{z-1}}{4\pi }.  \label{charge}
\end{equation}%
The electrostatic potential difference ($U$) between the horizon and
infinity is defined as%
\begin{equation}
U=A_{\mu }\chi ^{\mu }\left\vert _{r\rightarrow \infty }-A_{\mu }\chi ^{\mu
}\right\vert _{r=r_{+}},  \label{Pot}
\end{equation}%
Using Eqs. (\ref{at}) and (\ref{Pot}), one can obtain the electric potential%
\begin{equation}
U=\frac{qb^{2z-2}}{(n+z-3)r_{+}^{n+z-3}}\mathbf{F}\left( \frac{1}{2},\frac{%
n+z-3}{2n-2},\frac{3n+z-5}{2n-2},1-\Upsilon _{+}^{2}\right) ,
\label{elecpot}
\end{equation}%
which is the same for both $z=n+1$\ and $z\neq n+1$\ cases. In order to
investigate the first law of black hole thermodynamics, we should obtain the
Smarr-type formula for mass (\ref{Mass}). With Eqs. (\ref{mr+}), (\ref%
{charge}) and (\ref{entropy}) at hand, the mass can be written as a function
of extensive thermodynamic quantities $S$ and $Q$ in the form of%
\begin{equation}
M\left( S,Q\right) {=}\left\{ 
\begin{array}{ll}
\begin{array}{l}
\frac{(n-1)\left( 4S\right) ^{(n+z-1)/(n-1)}}{16\pi l^{z+1}}+\frac{\left(
n-1\right) (n-2)^{2}(4S)^{(n+z-3)/(n-1)}}{16\pi l^{z-1}\left( n+z-3\right)
^{2}}+\frac{\beta ^{2}\left( 4S\right) ^{(n-z+1)/(n-1)}(\text{{\ }}1-\Gamma )%
}{4\pi {l}^{z-1}{b}^{2(1-\,z)}(n-z+1)} \\ 
+\frac{4(n-1)\pi Q^{2}{b}^{2z-2}l^{1-z}\Gamma }{(n+z-3)(n-z+1)\left(
4S\right) ^{\left( n+z-3\right) /(n-1)}}{}\mathbf{F}\left( 1,\frac{2n+z-4}{%
2n-2},\frac{3n+z-5}{2n-2},1-\Gamma ^{2}\right) ,%
\end{array}
& \text{for }z\neq n+1, \\ 
&  \\ 
\frac{(n-1)\left( 4S\right) ^{2n/(n-1)}}{16\pi l^{n+2}}+\frac{(n-2)^{2}S^{2}%
}{4\pi (n-1)l^{n}}-\frac{\beta ^{2}b^{2n}}{4\pi (n-1)l^{n}}\left[ 1-\Gamma
+\ln \left( \frac{1+\Gamma }{2}\right) \right] , & \text{for }z=n+1,%
\end{array}%
\right.  \label{M}
\end{equation}%
%
%
%
%
%
where $\Gamma =\sqrt{1+\pi ^{2}Q^{2}/(\beta ^{2}S^{2})}$. Calculations show
that intensive quantities%
\begin{equation}
T=\left( \frac{\partial M}{\partial S}\right) _{Q}\text{ \ \ \ \ and \ \ \ \ 
}U=\left( \frac{\partial M}{\partial Q}\right) _{S},  \label{intqua}
\end{equation}%
coincide with those computed by Eqs. (\ref{Temp}) and (\ref{elecpot}).
Therefore, the thermodynamics quantities satisfy the first law of
thermodynamics%
\begin{equation}
dM=TdS+UdQ,  \label{TFL}
\end{equation}%
for both solutions for $z=n+1$\ and $z\neq n+1$.

In next part of this section, we will discuss the Hawking-Page and inside
black hole phase transitions for our Lifshitz solutions.

\subsection{Black hole phase transitions}

\subsubsection{Hawking-Page phase transition}

\begin{figure*}[t]
\centering{%
\subfigure[~$U=0.6$ (Linearly Charged Case)]{
   \label{fighp1a}\includegraphics[width=.46\textwidth]{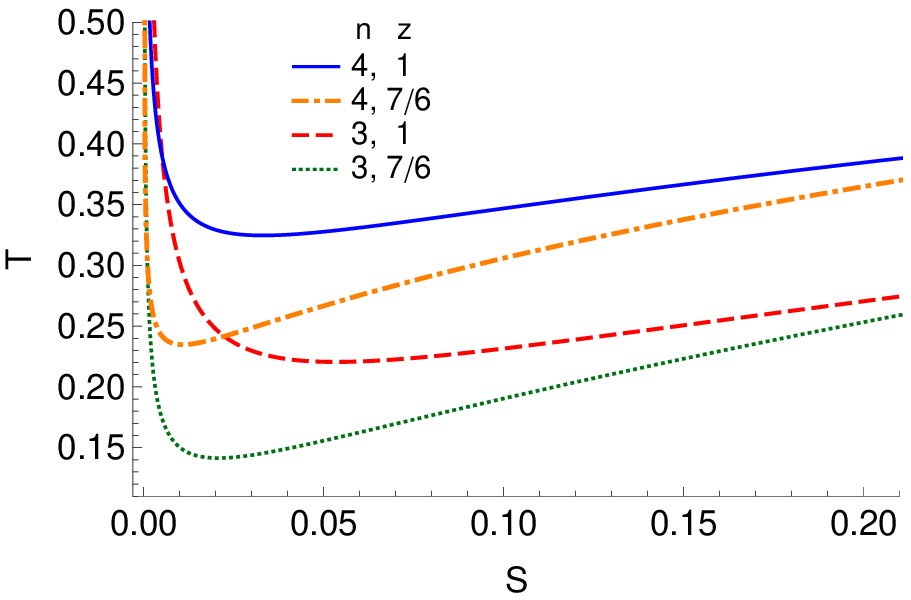}\qquad}} 
\subfigure[~$U=0.3$, $\beta=5$ (Nonlinearly Charged Case)]{
   \label{fighp1b}\includegraphics[width=.46\textwidth]{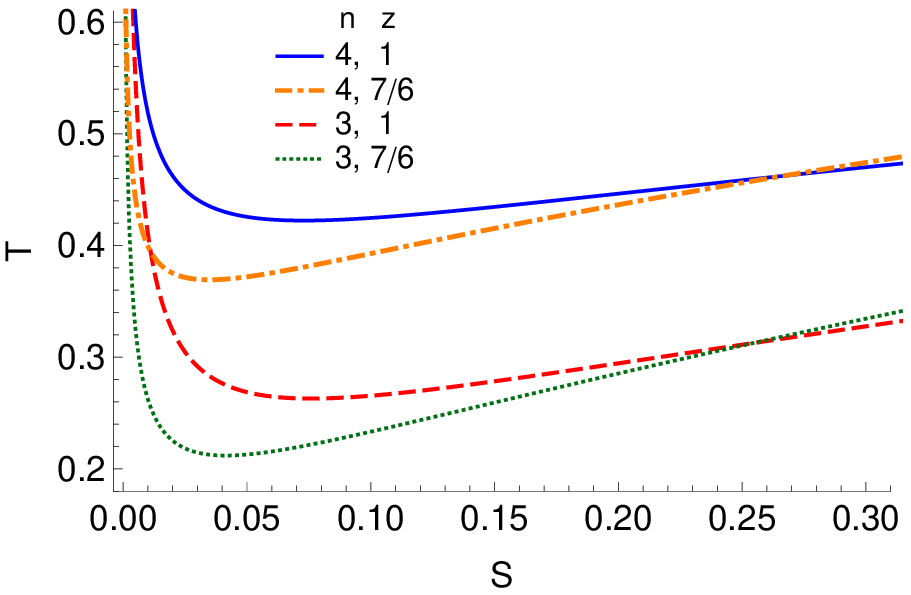}}
\caption{The behavior of $T$ versus $S$ for $l=1$, $b=1$ .}
\label{hp1}
\end{figure*}

\begin{figure*}[t]
\centering{%
\subfigure[~$z=1$]{
   \label{fighp2a}\includegraphics[width=.46\textwidth]{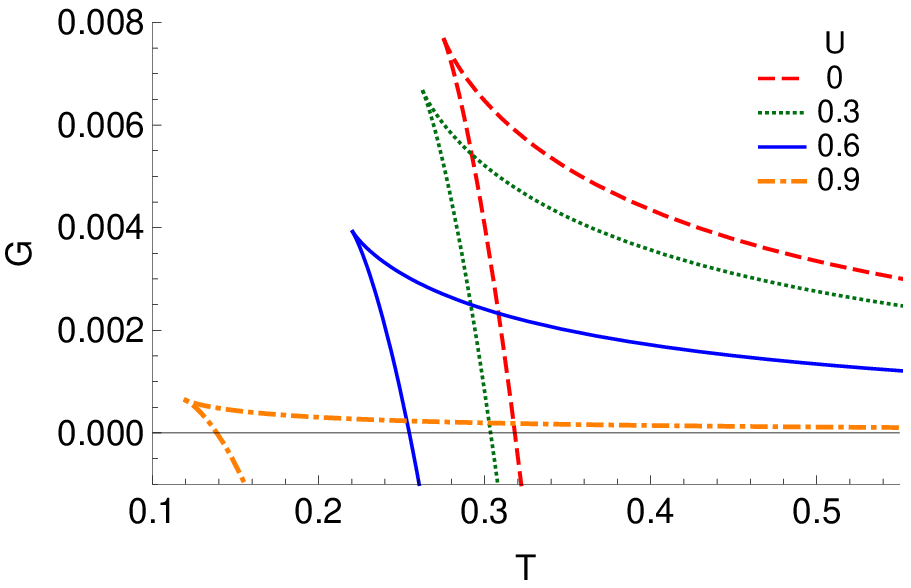}\qquad}} 
\subfigure[~$U=0.3$]{
   \label{fighp2b}\includegraphics[width=.46\textwidth]{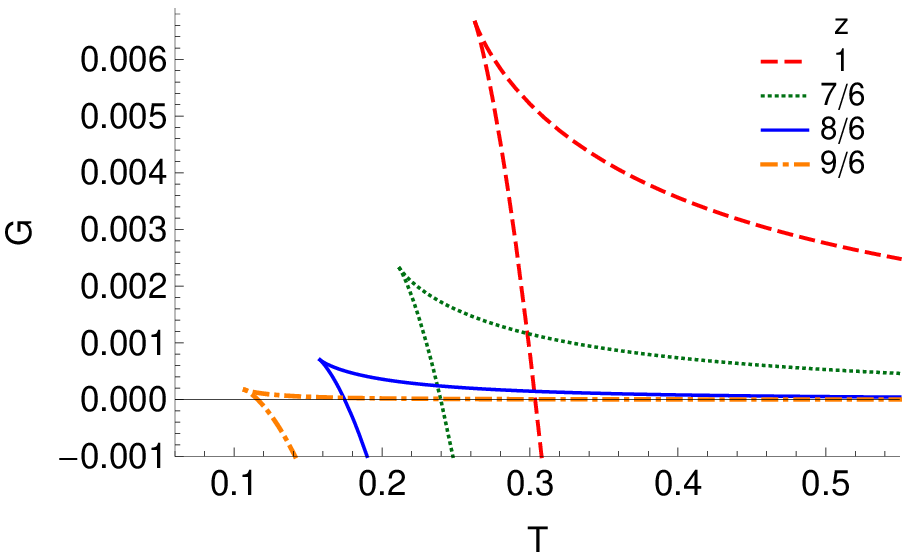}}
\caption{The behavior of $G$ versus $T$ for linearly charged case with $l=1$%
, $b=1$ and $n=3$. }
\label{hp2}
\end{figure*}

\begin{figure*}[t]
\centering{%
\subfigure[~$U=0.6$, $z=1$]{
   \label{fighp3a}\includegraphics[width=.46\textwidth]{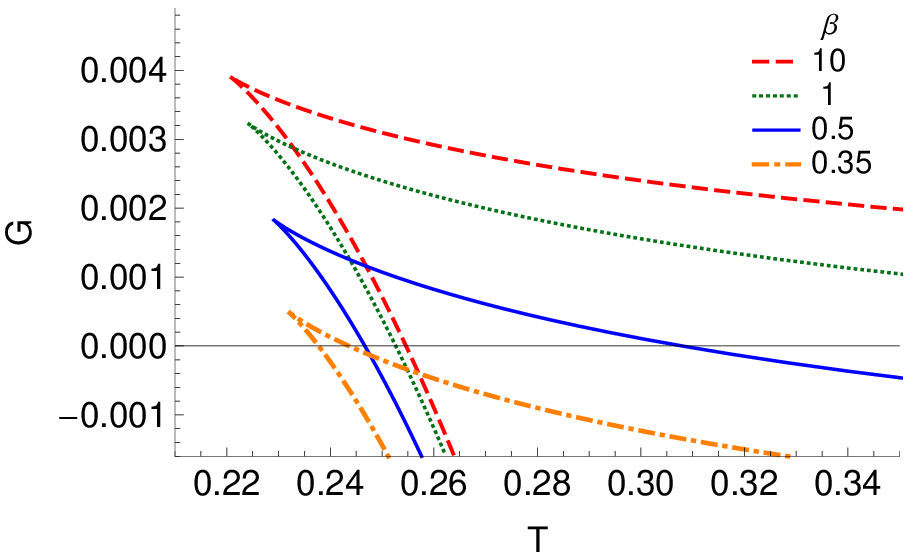}\qquad}} 
\subfigure[~$U=0.3$, $\beta=5$]{
   \label{fighp3b}\includegraphics[width=.46\textwidth]{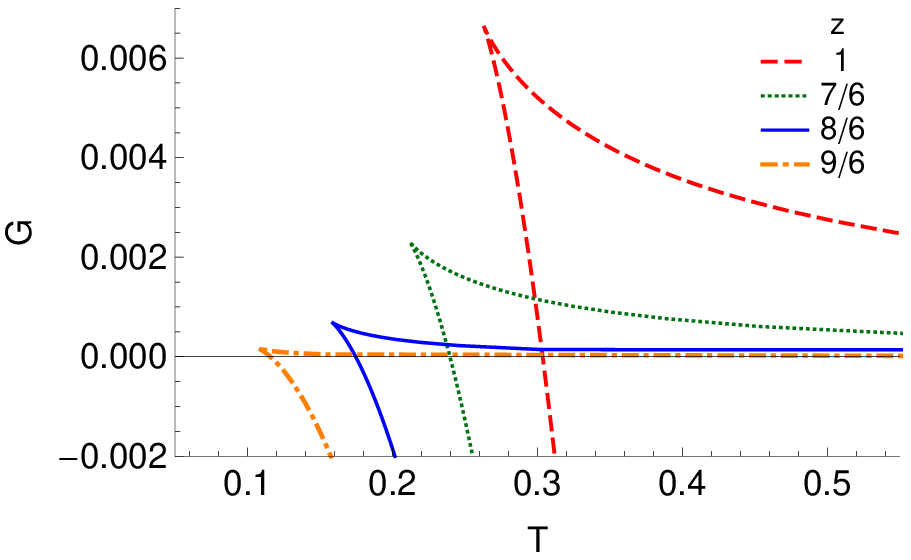}}
\caption{The behavior of $G$ versus $T$ for nonlinearly charged case with $%
l=1$, $b=1$ and $n=3$.}
\label{hp3}
\end{figure*}

As it is clear from Fig. \ref{fig1}, there are some parameter choices for
which we have extreme black holes and therefore zero temperature. In
addition, as one can see from Fig. \ref{hp1}, there are some other choices
of parameters that show a non-zero positive minimum for temperature $T_{\min
}$. The influences of different parameters on $T_{\min }$ can be seen from
Fig. \ref{hp1}. When we increase the dimension $n$, $T_{\min }$ increases
too, while it decreases with increasing $z$. Comparing Figs. \ref{fighp1a}
and \ref{fighp1b}, one finds out that the effect of nonlinearity implies
increasing in $T_{\min }$. The behaviors illustrated in Fig. \ref{hp1}
present a Hawking-Page phase transition for the obtained solutions. Let us
have a closer look on Fig. \ref{hp1}. In the first part of $T-S$ curves
where we have small black holes (note that $S=r_{+}^{n-1}/4$), $\partial
T/\partial S<0$ which implies negative heat capacity and therefore small
black holes are thermally unstable. But, in the large black holes part of
the curves we have a positive heat capacity and therefore large black holes
are thermally stable. In addition to small and large black holes, we have a
thermal Lifshitz or radiation solution too. Since the small black holes are
thermally unstable, system has two choices between large black hole and
thermal Lifshitz that chooses to be on one of them according to the Gibbs
free energy. The Gibbs free energy%
\begin{equation}
G\left( T,U\right) =M-TS-QU,
\end{equation}%
can be obtained by using (\ref{Temp}), (\ref{entropy}), (\ref{Mass}), (\ref%
{charge}) and (\ref{elecpot}). Figs. \ref{hp2} and \ref{hp3} show the
behavior of Gibbs free energy for some choices of parameters. The two up and
bottom branches correspond to small and large black holes, respectively. The
positive Gibbs free energy shows that the system is in radiation phase while
there is a Hawking-Page phase transition at intersection point of bottom
branch and $G=0$. This fact that the Gibbs free energy of large black holes
always have the lower energy in comparison to small ones confirms the above
arguments about the thermal stability of them. As one moves rightward on
temperature axis in $G-T$ diagram, first experiences radiation regime or
thermal Lifshitz solution for which $G>0$. At $G=0$, the Hawking-Page phase
transition between thermal Lifshitz and large black holes occurs and for $%
G<0 $, we are at large black hole phase. The temperature at which phase
transition occurs is called Hawking-Page temperature $T_{HP}$. Effects of
change in electric potential $U$, critical exponent $z$ and nonlinearity
parameter $\beta $ can be seen from Figs. \ref{hp2} and \ref{hp3}. Increase
in electric potential $U$ and critical exponent $z$ makes $T_{HP}$ lower.
Also, the lower the nonlinearity parameter $\beta $ is, the lower
Hawking-Page temperature $T_{HP}$ is. Note that lower $\beta $ makes the
electrodynamics more affected by nonlinearity. 

\subsubsection{Phase transitions inside the black hole\label{insphase}}

There are at least three well-known ways to discuss the phase transitions
inside the black hole. Two of these ways are based on macroscopic point of
view and one of them is based on microscopic viewpoint. The two macroscopic
ways are Davies \cite{Davies1} and Landau-Lifshitz \cite{LL1,LL2} methods
that discuss, respectively, the behavior of heat capacities and
thermodynamic fluctuations. Thermodynamic geometry or Ruppeiner geometry 
\cite{Rup2,Rup5,Rup6} is the microscopic way which discusses the phase
transitions in addition to type and strength of interactions. In what
follows, we discuss the relation between the phase transitions predicted by
Ruppeiner geometry and Davies method. Next, we will turn to Landau-Lifshitz
theory of thermodynamic fluctuations.

\subsubsection*{Ruppeiner and Davies phase transitions}

In order to discuss thermodynamic geometry, one should study the
divergences, sign and magnitude of Ricci scalar corresponding to Ruppeiner
metric (usually called Ruppeiner invariant) to determine phase transitions
and strength and type of dominated interaction between possible black hole
molecules \cite{Rup2,Rup5,Rup6}. To do that, we define the Ruppeiner metric
in ($M$, $Q$) space where the entropy $S$ is thermodynamic potential as%
\begin{equation}
g_{\alpha \beta }=-\frac{\partial ^{2}S}{\partial X^{\alpha }\partial
X^{\beta }},\text{ \ \ \ \ }X^{\alpha }=(M,Q).
\end{equation}%
The above metric can also be rewritten in the Weinhold form%
\begin{equation}
g_{\alpha \beta }=\frac{1}{T}\frac{\partial ^{2}M}{\partial Y^{\alpha
}\partial Y^{\beta }},\text{ \ \ \ \ }Y^{\alpha }=(S,Q).  \label{MetGTD}
\end{equation}%
The Ruppeiner invariant corresponding to (\ref{MetGTD}) can be expressed in
a general form as%
\begin{equation}
\mathfrak{R}=\frac{\mathfrak{N}(S,Q)}{\mathfrak{D}(S,Q)},  \label{RupMax}
\end{equation}%
where $\mathfrak{N}$ and $\mathfrak{D}$ stand for numerator and denominator
of $\mathfrak{R}$. The divergences of Ruppeiner invariant is determined by
roots of $\mathfrak{D}$ which is equal to $T\left[ \mathbf{H}_{S,Q}^{M}%
\right] ^{2}$ where $\mathbf{H}_{S,Q}^{M}=M_{SS}M_{QQ}-M_{SQ}^{2}$ is
determinant of Hessian matrix and $X_{YZ}=\partial ^{2}X/\partial Y\partial
Z $. Of course, at these divergence points the numerator $\mathfrak{N}$\
should be finite. These divergences show both zero temperature and vanishing 
$\mathbf{H}_{S,Q}^{M}$. The root of $\mathbf{H}_{S,Q}^{M}$ may show us the
boundary between thermal stability and instability. For thermal stability,
in addition to positivity of determinant of Hessian matrix,\ $M_{QQ}$ and $%
M_{SS}$ should be positive too \cite{stability1,stability2}.

It is remarkable to note that at the point where $M_{SS}$ vanishes or
equivalently heat capacity at constant charge $C_{Q}$ diverges, we have a
thermally unstable system due to negativity of $\mathbf{H}_{S,Q}^{M}$ if $%
M_{SQ}\neq 0$ (which occurs in many of black hole systems). Thus, the heat
capacity at constant charge $C_{Q}$ cannot be suitable thermodynamic
quantity to show phase transition of such systems when we have two changing
thermodynamic parameters, for instance $S$ and $Q$. There are some works in
literature (for instance \cite{hendi}) in which the correctness of Ruppeiner
method for recognizing the phase transitions has been judged by comparing
the Ruppeiner and $C_{Q}$ transition points. This procedure is of course
seems to be incorrect according to what we pointed out above. Also, as we
discussed in the introduction, divergences of $\mathfrak{R}$ are safer in
order to determine phase transitions. On the other hand, in \cite{RupGeo1}
and \cite{RupGeo5}, authors have suggested some suitable thermodynamic
quantities to show the phase transitions predicted by Ruppeiner invariant.
These quantities are specific heat at constant electrical potential, $C_{U}$%
, analog of volume expansion coefficient, $\alpha $, and analog of
isothermal compressibility coefficient $\kappa _{T}$ defined as%
\begin{equation}
C_{U}=T\left( \frac{\partial S}{\partial T}\right) _{U}\text{, \ }\alpha =%
\frac{1}{Q}\left( \frac{\partial Q}{\partial T}\right) _{U},\ \ \ \kappa
_{T}=\frac{1}{Q}\left( \frac{\partial Q}{\partial U}\right) _{T}\text{.}
\end{equation}%
As one can see in appendix \ref{app1}, these thermodynamic quantities have
the forms%
\begin{equation}
C_{U}=T\frac{M_{SS}}{\mathbf{H}_{S,Q}^{M}},\text{ \ \ \ \ }\alpha =-\frac{1}{%
Q}\frac{M_{SQ}}{\mathbf{H}_{S,Q}^{M}},\ \ \ \kappa _{T}=-\alpha \left. \frac{%
\partial T}{\partial U}\right\vert _{Q}.
\end{equation}%
It is obvious that these quantities show the same phase transitions as the
Ruppeiner geometry because all of them diverge at roots of $\mathbf{H}%
_{S,Q}^{M}$ and $C_{U}$ vanishes at zero temperature where $\mathfrak{R}$
diverges. To show the coincidence of Ruppeiner phase transitions and $C_{U}$
divergences, some proofs have been presented in \cite{RupGeo61,RupGeo10}.
The above quantities can be considered as improved Davies quantities \cite%
{Davies1} which present the phase transitions coincided with Ruppeiner ones.
In the next part, we study the Landau-Lifshitz theory of thermodynamic
fluctuations to explore the possible signature of black hole phase
transitions on properties of black hole radiance.

\subsubsection*{Landau-Lifshitz theory (nonextreme/extreme phase transition)}

Here, we seek for any possible effect of transition on black hole radiance
by using Landau-Lifshitz theory of thermodynamic fluctuations \cite{LL1,LL2}%
. We focus on ($3+1$)-dimensional linearly charged case. The extension to
higher dimensional or nonlinearly charged cases is trivial and give no novel
result. Based on Landau-Lifshitz theory \cite{LL1,LL2}, in a
fluctuation-dissipative process, the flux $\dot{X}_{i}$ of a given
thermodynamic quantity $X_{i}$ is given by%
\begin{equation}
\dot{X}_{i}=-\sum_{j}\Gamma _{ij}\chi _{j},
\end{equation}%
where dot shows temporal derivative and $\chi _{i}$ and $\Gamma _{ij}$ are
respectively the thermodynamic force conjugate to the flux $\dot{X}_{i}$ and
the phenomenological transport coefficients. In addition, the rate of
entropy production is expressed by%
\begin{equation}
\dot{S}=\sum_{i}\pm \chi _{i}\dot{X}_{i},
\end{equation}%
where "$+$" ("$-$") holds for the entropy rate contributions which come from
the non-concave (concave) parts of $S$. The second moments corresponding to
the fluxes' fluctuations are (we set $k_{B}=1$)%
\begin{equation}
\left\langle \delta \dot{X}_{i}\delta \dot{X}_{j}\right\rangle =\left(
\Gamma _{ij}+\Gamma _{ji}\right) \delta _{ij},  \label{smom}
\end{equation}%
where the mean value with respect to the steady state is denoted by the
angular brackets and the fluctuations $\delta \dot{X}_{i}$ are the
spontaneous deviations from the value of steady state $\left\langle \dot{X}%
_{i}\right\rangle $. To guarantee that correlations are zero when two fluxes
are independent, the Kronecker $\delta _{ij}$ is put in Eq. (\ref{smom}).

According to \cite{Kord2}, the mass $M$, electric potential energy $U$ and
temperature $T$ can be obtained for ($3+1$)-dimensional linearly charged
case as%
\begin{equation}
M=\frac{(4S)^{(z+2)/2}}{8\pi l^{z+1}}+\frac{(4S)^{z/2}}{8\pi z^{2}l^{z-1}}+%
\frac{2\pi Q^{2}b^{2z-2}}{zl^{z-1}(4S)^{z/2}},  \label{sm}
\end{equation}%
\begin{equation}
U=\frac{\pi b^{2z-2}Q}{z2^{z-2}l^{z-1}S^{z/2}}\text{ and }T=\frac{2^{z-4}\Xi 
}{\pi zl^{z-1}S^{z/2+1}},  \label{TTT}
\end{equation}%
where%
\begin{equation*}
\Xi =S^{z}+4z(z+2)S^{z+1}l^{-2}-4^{2-z}z\pi ^{2}b^{2z-2}Q^{2}.
\end{equation*}%
We know that in extreme black hole case, the Hawking temperature on the
event horizon vanishes and therefore in this case we have $\Xi =0$. Using
Eq. (\ref{sm}), we can obtain the entropy production rate as%
\begin{equation}
\dot{S}(M,Q)=\chi _{M}\dot{M}-\chi _{Q}\dot{Q},
\end{equation}%
where%
\begin{equation*}
\chi _{M}=\frac{\pi zl^{z-1}S^{z/2+1}}{2^{z-4}\Xi }\text{ and }\chi _{Q}=%
\frac{\pi ^{2}b^{2z-2}QS}{4^{z-2}\Xi }.
\end{equation*}%
The mass loss rate is given by \cite{His}%
\begin{equation}
\frac{dM}{dt}=-b\alpha \sigma T^{4}+U\frac{dQ}{dt}.  \label{Md}
\end{equation}%
The first term on the right side of Eq. (\ref{Md}) is the thermal mass loss
corresponding to Hawking radiation which is just the Stefan-Boltzmann law,
with $b=\pi ^{2}/15$ (we set $\hbar =1$) as the radiation constant. The
constant $\alpha $ depends on the number of species of massless particles
and the quantity $\sigma $ is the cross-section of geometrical optics. The
second term on the right side of Eq. (\ref{Md}) is responsible for the loss
of mass corresponding to charged particles. In fact, it is $UdQ$ term which
rises in first law of black hole mechanics.

With references to what explained and computed above, one can calculate the
second moments or correlation functions of the thermodynamical quantities%
\begin{equation}
\left\langle \delta \dot{M}\delta \dot{M}\right\rangle =-\frac{2^{z-3}\Xi }{%
\pi zl^{z-1}S^{z/2+1}}\dot{M},\text{ \ \ \ \ }\left\langle \delta \dot{Q}%
\delta \dot{Q}\right\rangle =\frac{2^{2z-5}b^{2-2z}\Xi }{\pi ^{2}SQ}\dot{Q},%
\text{ \ \ \ \ }\left\langle \delta \dot{M}\delta \dot{Q}\right\rangle
=U\left\langle \delta \dot{Q}\delta \dot{Q}\right\rangle ,
\end{equation}%
\begin{eqnarray}
\left\langle \delta \dot{S}\delta \dot{S}\right\rangle &=&\frac{\pi
^{2}z^{2}l^{2z-2}S^{z+2}}{4^{z-4}\Xi ^{2}}\left[ \left\langle \delta \dot{M}%
\delta \dot{M}\right\rangle +\frac{\pi ^{2}b^{4z-4}Q^{2}}{%
4^{z-2}z^{2}l^{2z-2}S^{z}}\left\langle \delta \dot{Q}\delta \dot{Q}%
\right\rangle -\frac{\pi b^{2z-2}Q}{2^{z-3}zl^{z-1}S^{z/2}}\left\langle
\delta \dot{M}\delta \dot{Q}\right\rangle \right]  \notag \\
&=&-\frac{\pi zl^{z-1}S^{z/2+1}}{2^{z-5}\Xi }\left[ \dot{M}+\frac{\pi
b^{2z-2}Q}{z2^{z-2}l^{z-1}S^{z/2}}\dot{Q}\right]
\end{eqnarray}%
%
%
%
%
%
{\footnotesize 
\begin{eqnarray}
\left\langle \delta \dot{T}\delta \dot{T}\right\rangle &=&\frac{\left[
(z-2)S^{z}+4z^{2}(z+2)l^{-2}S^{z+1}+\pi ^{2}z(z+2)4^{2-z}b^{2z-2}Q^{2}\right]
^{2}}{4S^{2}\Xi ^{2}}\left\langle \delta \dot{M}\delta \dot{M}\right\rangle 
\notag \\
&&+\frac{4^{2-z}\pi ^{2}b^{4z-4}Q^{2}\left[
(z-1)S^{z}+4z^{2}(z+2)l^{-2}S^{z+1}+\pi ^{2}z4^{2-z}b^{2z-2}Q^{2}\right] ^{2}%
}{z^{2}l^{2z-2}S^{z+2}\Xi ^{2}}\left\langle \delta \dot{Q}\delta \dot{Q}%
\right\rangle  \notag \\
&&-\frac{\pi b^{2z-2}Q\left[ (z-1)S^{z}+4z^{2}(z+2)l^{-2}S^{z+1}+\pi
^{2}z4^{2-z}b^{2z-2}Q^{2}\right] \left[ (z-2)S^{z}+4z^{2}(z+2)l^{-2}S^{z+1}+%
\pi ^{2}z(z+2)4^{2-z}b^{2z-2}Q^{2}\right] }{z2^{z-2}l^{z-1}S^{z/2+2}\Xi ^{2}}%
\left\langle \delta \dot{M}\delta \dot{Q}\right\rangle  \notag \\
&=&-\frac{2^{z-5}}{\pi zl^{z-1}S^{z/2+3}\Xi }\left\{ \left[
(z-2)S^{z}+4z^{2}(z+2)l^{-2}S^{z+1}+\pi ^{2}z(z+2)4^{2-z}b^{2z-2}Q^{2}\right]
^{2}\dot{M}\right.  \notag \\
&&\left. -\frac{\pi b^{2z-2}Q}{2^{z-4}zl^{z-1}S^{z/2}}\left[
(z-1)S^{z}+4z^{2}(z+2)l^{-2}S^{z+1}+\pi ^{2}z4^{2-z}b^{2z-2}Q^{2}\right] %
\left[ S^{z}-16\pi ^{2}4^{-z}Q^{2}b^{2z-2}z(z+1)\right] \dot{Q}\right\}
\end{eqnarray}%
} 
\begin{eqnarray}
\left\langle \delta \dot{S}\delta \dot{T}\right\rangle &=&\frac{\pi zS^{z/2}%
\left[ (z-2)S^{z}+4z^{2}(z+2)l^{-2}S^{z+1}+\pi ^{2}z(z+2)4^{2-z}b^{2z-2}Q^{2}%
\right] }{2^{z-3}l^{z-1}\Xi ^{2}}\left\langle \delta \dot{M}\delta \dot{M}%
\right\rangle  \notag \\
&&+\frac{\pi ^{3}b^{4z-4}Q^{2}\left[ (z-1)S^{z}+4z^{2}(z+2)l^{-2}S^{z+1}+\pi
^{2}z4^{2-z}b^{2z-2}Q^{2}\right] }{z2^{3z-8}l^{z-1}S^{z/2}\Xi ^{2}}%
\left\langle \delta \dot{Q}\delta \dot{Q}\right\rangle  \notag \\
&&-\frac{\pi ^{2}b^{2z-2}Q\left[ (3z-4)S^{z}+12z^{2}(z+2)l^{-2}S^{z+1}+\pi
^{2}z(z+4)4^{2-z}b^{2z-2}Q^{2}\right] }{2^{2z-5}\Xi ^{2}}\left\langle \delta 
\dot{M}\delta \dot{Q}\right\rangle  \notag \\
&=&-\frac{\left[ (z-2)S^{z}+4z^{2}(z+2)l^{-2}S^{z+1}+\pi
^{2}z(z+2)4^{2-z}b^{2z-2}Q^{2}\right] }{l^{2z-2}S\Xi }\left[ \dot{M}+\frac{%
\pi b^{2z-2}l^{z-1}Q}{2^{z-2}zS^{\frac{1}{2}z}}\dot{Q}\right] .  \notag \\
&&
\end{eqnarray}%
It is clear that second moments $\left\langle \delta \dot{S}\delta \dot{S}%
\right\rangle $, $\left\langle \delta \dot{T}\delta \dot{T}\right\rangle $
and $\left\langle \delta \dot{S}\delta \dot{T}\right\rangle $ diverge for
extreme black hole case where $\Xi $ vanishes (see Eq. (\ref{TTT})). It
means that there is a phase transition in this case. This phase transition
is between extreme and nonextreme black holes for which we have a sudden
change in emission properties. In nonextreme case, the black hole can give
off particles and radiation through both spontaneous Hawking emission and
superradiant scattering whereas in extreme case, the black hole can just
radiate via superradiant scattering.

As one can see from Eq. (\ref{Md}), $\dot{M}$ and $\dot{Q}$ are related.
Therefore, all of the above second moments can be reexpressed in terms of $%
\dot{Q}$. Let us calculate $\dot{Q}$ for our case. The rate of charge loss
can be stated as%
\begin{equation}
-\frac{dQ}{dt}=e\int_{r_{+}}^{\infty }\int_{0}^{2\pi }\int_{0}^{\pi }\sqrt{-g%
}\Gamma d\theta d\phi dr,  \label{dQdt}
\end{equation}%
where $\Gamma $ is the rate of electron-positron pair creation per
four-volume and $e$ is charge of electron. According to Schwinger's theory 
\cite{Sch} for ($3+1$)-dimensions, the rate of electron-positron pair
creation in a constant electric field $E$ is%
\begin{equation}
\Gamma =\frac{4e^{2}b^{4z-4}}{\pi l^{2z-2}}E^{2}\exp \left( -\frac{1}{EQ_{0}}%
\right) \left[ 1+O\left( \frac{e^{3}E}{m^{2}}\right) +\cdots \right] ,
\label{Gam1}
\end{equation}%
where $Q_{0}=4\pi eb^{2z-2}/\pi m^{2}l^{z-1}$ and $m$ is the mass of
electron. In the presence of linear Maxwell electrodynamics, the electric
field is $E=Q/r^{z+1}$ and therefore%
\begin{equation}
\Gamma =\frac{4e^{2}b^{4z-4}Q^{2}}{\pi l^{2z-2}r^{2z+2}}\exp \left( -\frac{%
r^{z+1}}{QQ_{0}}\right) \left[ 1+O\left( \frac{e^{3}Q}{m^{2}r^{z+1}}\right)
+\cdots \right] .  \label{Gam2}
\end{equation}%
Combining Eqs. (\ref{Gam1}) and (\ref{Gam2}), we arrive at 
\begin{equation}
\frac{dQ}{dt}=-\frac{16e^{3}b^{4z-4}Q^{(z+2)/(z+1)}}{%
(z+1)l^{3z-3}Q_{0}^{z/(z+1)}}\Gamma \left[ -\frac{z}{z+1},\frac{r_{+}^{z+1}}{%
QQ_{0}}\right] ,  \label{dQ2}
\end{equation}%
where $\Gamma \left[ a,b\right] $ is incomplete gamma function. When $%
r_{+}\gg Q$, Eq. (\ref{dQ2}) reduces to 
\begin{equation}
\frac{dQ}{dt}\approx -\frac{64e^{4}b^{6z-6}Q^{3}}{%
(z+1)m^{2}l^{4z-4}r_{+}^{2z+1}}\exp \left( -\frac{r_{+}^{z+1}}{QQ_{0}}%
\right) +\cdots ,
\end{equation}%
where we have used%
\begin{equation}
\Gamma \left[ -\frac{z}{z+1},x\right] \approx \exp \left( -x\right)
x^{1/(z+1)}\left[ \frac{1}{x^{2}}+O\left( \frac{1}{x^{3}}\right) +\cdots %
\right] ,
\end{equation}%
in which $x^{-1}\ll 1$.

In the following section, we turn to study thermodynamic geometry of our
black hole solutions to figure out the behavior of black hole possible
molecules and phase transitions.

\section{Ruppeiner geometry\label{RG}}

In this section we study thermodynamic geometry of the Lifshitz-dilaton
black holes for linearly Maxwell and nonlinearly BI gauge fields,
separately. We have introduced this method in subsection \ref{insphase} with
focus on the study of the phase transitions which occur at divergence of
Ruppeiner invariant $\mathfrak{R}$. In addition to divergences, $\mathfrak{R}
$ has other properties which give us information about thermodynamic of the
system. The sign of $\mathfrak{R}$ gives us the information about the
dominated interaction between possible black hole molecules while its
magnitude measures the average number of correlated Planck areas on the
event horizon \cite{Rup2,Rup05,Rup5,Rup6}. $\mathfrak{R}>0$ means the
domination of repulsive interaction, $\mathfrak{R}<0$ shows the attraction
dominated regime and when\textbf{\ }$\mathfrak{R}$ vanishes the system
behaves like ideal gas i.e. there is no interaction. In continue, we first
study thermodynamic geometry in the presence of linear Maxwell
electrodynamics. Then, we extend our study to nonlinearly charged black
holes where BI electrodynamics has been employed. There is just a necessary
comment. As we stated before in subsection \ref{insphase}, for thermal
stability, $M_{QQ}$, $M_{SS}$ and $\mathbf{H}%
_{S,Q}^{M}=M_{SS}M_{QQ}-M_{SQ}^{2}$ should be positive \cite%
{stability1,stability2}. One can show that the positivity of $\mathbf{H}%
_{S,Q}^{M}$ and\textbf{\ }$M_{QQ}$ ($M_{SS}$) imposes the positivity of $%
M_{SS}$ ($M_{QQ}$). Therefore, we just turn to study the signs of $\mathbf{H}%
_{S,Q}^{M}$ and\textbf{\ }$M_{QQ}$ in our following discussions to guarantee
the thermal stability.

\subsection{Linear Maxwell case}

The mass and Hawking temperature of black holes in the presence of linear
Maxwell (LM) electrodynamics are 
\begin{equation}
T_{LM}=\frac{(n+z-1)r_{+}^{z}}{4\pi l^{z+1}}+\allowbreak \frac{%
(n-2)^{2}l^{1-z}}{4\pi (n+z-3)r_{+}^{2-z}}-\frac{q^{2}l^{z-1}b^{2z-2}}{2\pi
(n-1)r_{+}^{2n+z-4}},  \label{Tmax}
\end{equation}%
%
%
%
%
%
\begin{equation}
M_{LM}(S,Q)=\frac{(n-1)(4S)^{(n+z-1)/(n-1)}}{16\pi l^{z+1}}+\frac{%
(n-1)(n-2)^{2}(4S)^{(n+z-3)/(n-1)}}{16\pi (n+z-3)^{2}l^{z-1}}+\frac{2\pi
Q^{2}b^{2z-2}(4S)^{(3-n-z)/(n-1)}}{(n+z-3)l^{z-1}}.  \label{Mmax}
\end{equation}%
%
%
%
%
%
As we mentioned above, for investigating thermal stability we need to check
the signs of $M_{QQ}$ and $\mathbf{H}_{S,Q}^{M}$. In our case 
\begin{equation}
M_{QQ}=\frac{\pi b^{2z-2}S^{-\left( n+z-3\right) /\left( n-1\right) }}{%
\left( n+z-3\right) l^{z-1}2^{2\left( z-2\right) /\left( n-1\right) }}>0,
\label{MQQ}
\end{equation}%
Thus, in order to disclose the thermal stability of system, we need to study
the sign of determinant of Hessian matrix. We find 
\begin{equation}
\mathbf{H}_{S,Q}^{M}=\left\{ 
\begin{tabular}{lc}
$\frac{(z-2)(n-2)^{2}b^{2z-2}S^{-2\left[ 2+(z-2)/(n-1)\right] }}{%
4(n-1)(n+z-3)^{2}l^{2z-2}}\mathfrak{F}(S,Q)$ & $z\neq 2$ \\ 
&  \\ 
$\frac{(n+1)b^{2}2^{(n-5)/(1-n)}S^{2(n-2)/(1-n)}}{(n-1)^{2}l^{4}}$ & $z=2$%
\end{tabular}%
\right. ,  \label{deno}
\end{equation}%
where 
\begin{equation}
\mathfrak{F}(S,Q)\equiv \left[ S^{2[1+(z-2)/(n-1)]}+\frac{%
z(n+z-3)(n+z-1)S^{2[1+(z-1)/(n-1)]}}{2^{-4/(n-1)}(z-2)(n-2)^{2}l^{2}}-\frac{%
\pi ^{2}(n+z-3)Q^{2}2^{(n-4z+7)/(n-1)}}{(n-1)(n-2)^{2}b^{2(1-z)}}\right] ,
\end{equation}%
%
%
%
%
%
The numerator $\mathfrak{N}$ of (\ref{RupMax}) is a complicated finite
function of $S$ and $Q$ in this case, including long terms that we do not
express it explicitly for economic reasons. However, as it was mentioned in
subsection \ref{insphase}, one can find the denominator $\mathfrak{D}$ in
the form of 
\begin{equation}
\mathfrak{D}(S,Q)=T_{LM}\left[ \mathbf{H}_{S,Q}^{M}\right] ^{2},
\label{deno0}
\end{equation}%
where $T_{LM}$ and $\mathbf{H}_{S,Q}^{M}$ have been give in (\ref{Tmax}) and
(\ref{deno}) respectively.

\begin{figure*}[t]
\centering{%
\subfigure[~$n=3$, $z=1$, $Q_c=0.023$]{
   \label{fig2a}\includegraphics[width=.46\textwidth]{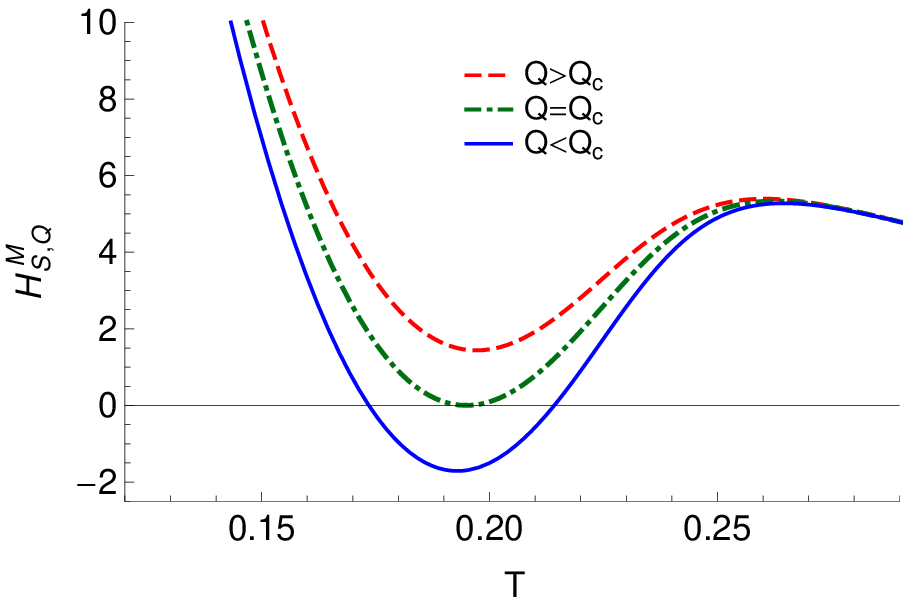}\qquad}} 
\subfigure[~$n=4$, $z=1.5$, $Q_c=0.003$]{\
\label{fig2b}\includegraphics[width=.46\textwidth]{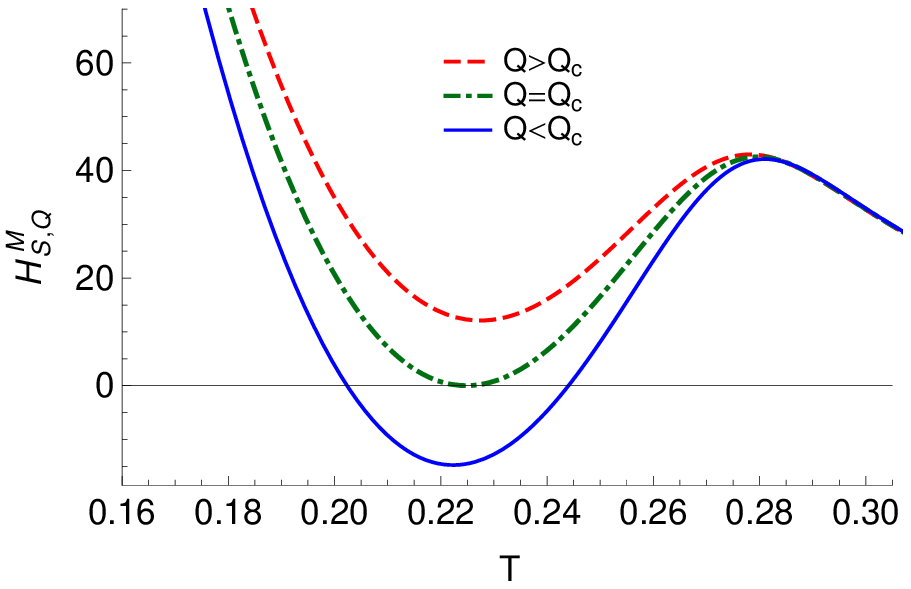}}
\caption{The behavior of $\mathbf{H}_{S,Q}^{M}$ versus $T$ for linear
Maxwell case with $l=b=1$.}
\label{fig2}
\end{figure*}

Having Eqs. (\ref{deno}) and (\ref{deno0}) at hand, we are in the position
to investigate the divergences of $\mathfrak{R}$, which play the central
role in thermodynamic geometry discussions and also thermal stability of
system. As one can see from Eqs. (\ref{deno}) and (\ref{deno0}), for $z=2$,
the divergences occur just in the case of the extremal black holes where $%
T_{LM}=0$. For $z\neq 2$, in addition to extremal black hole case, $%
\mathfrak{R}$ diverges in zeros of (\ref{deno}). In the latter case, we can
calculate the corresponding temperature by solving $\mathfrak{F}=0$ for $Q$
and then putting this $Q$ in Eq. (\ref{Tmax}) to arrive at 
\begin{equation}
\mathcal{T}=\frac{(n+z-1)2^{(2z-n+1)/(n-1)}S^{z/(n-1)}}{\pi (2-z)l^{z+1}}.
\label{TT}
\end{equation}%
The above temperature is negative for $z>2$ i.e. there is no black hole at
this diverging point and therefore the divergences of $\mathfrak{R}$ occur
just for extremal black hole case when $z>2$. However, for $z<2$ when $%
\mathcal{T}>0$, we can see an upper limit in entropy and charge of system.
The largest entropy $S$ for which $\mathfrak{F}=0$ (which we call it
critical entropy $S_{c}$) can be calculated by finding the extremum point
where $\partial \mathfrak{F/}\partial S=0$ as 
\begin{equation}
S_{c}^{2/(1-n)}=\frac{z(n+z-1)(n+z-2)2^{4/(n-1)}}{(2-z)(n-2)^{2}l^{2}},
\label{Sc}
\end{equation}%
at which 
\begin{equation}
Q_{c}^{2}=\frac{(n-2)^{2(n+z-2)}}{l^{-2(n+z-3)}b^{2(z-1)}\pi ^{2}}\frac{%
(n-1)(n+z-2)^{2-n-z}(n+z-1)^{3-n-z}}{2^{5}(n+z-3)}\left( \frac{2}{z}%
-1\right) ^{n+z-3},  \label{Qc}
\end{equation}%
and 
\begin{equation}
\left. \frac{\partial ^{2}\mathfrak{F}}{\partial S^{2}}\right\vert
_{S=S_{c}}=-\frac{\left( n+z-3\right) 4^{\left( n-2z+3\right) /\left(
n-1\right) }}{\left( n-1\right) ^{2}}\left[ \frac{\left( n+z-1\right) \left(
n+z-2\right) z}{\left( 2-z\right) \left( n-2\right) ^{2}l^{2}}\right]
^{2-z}<0.
\end{equation}%
One should note that the absolute value of $Q_{c}$ is also the largest
charge value which satisfies $\mathfrak{F}=0$. Another remark to be
mentioned is that (\ref{Sc}) imposes an upper limit on the size of black
hole too (see (\ref{entropy})). At this point, the corresponding temperature
can be obtained as 
\begin{equation}
T_{c}=\frac{(n-2)^{z}}{2\pi l\sqrt{\left( z(n+z-2)\right) ^{z}}}\left( \frac{%
n+z-1}{2-z}\right) ^{(2-z)/2}.  \label{Tc}
\end{equation}%
For charges greater than $Q_{c}$, the Ruppeiner invariant diverges only in
the case of extremal black holes. For $Q=Q_{c}$, in addition to $T_{LM}=0$,
we have one other divergence in $\mathfrak{R}$ specified by (\ref{Sc}) and (%
\ref{Tc}). For $Q<Q_{c}$, in addition to $T_{LM}=0$, we have at most two
other divergences since the order of polynomial in term of $S$ is always
lower than $3$ for $n\geq 3$ and $z<2$. One should note that, in latter
case, the temperature region between two divergences is not allowed since $%
\mathbf{H}_{S,Q}^{M}<0$ (Fig. (\ref{fig2})). 
\begin{figure*}[t]
\centering{%
\subfigure[~$n=3$, $z=1$, $Q_c=0.023$]{
   \label{fig3a}\includegraphics[width=.46\textwidth]{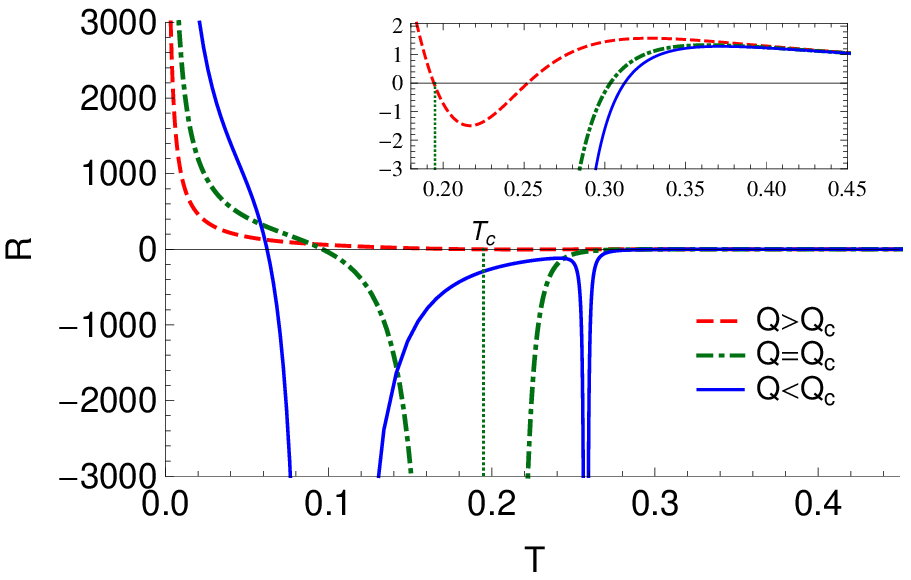}\qquad}} 
\subfigure[~$n=4$, $z=1.5$, $Q_c=0.003$]{
   \label{fig3b}\includegraphics[width=.46\textwidth]{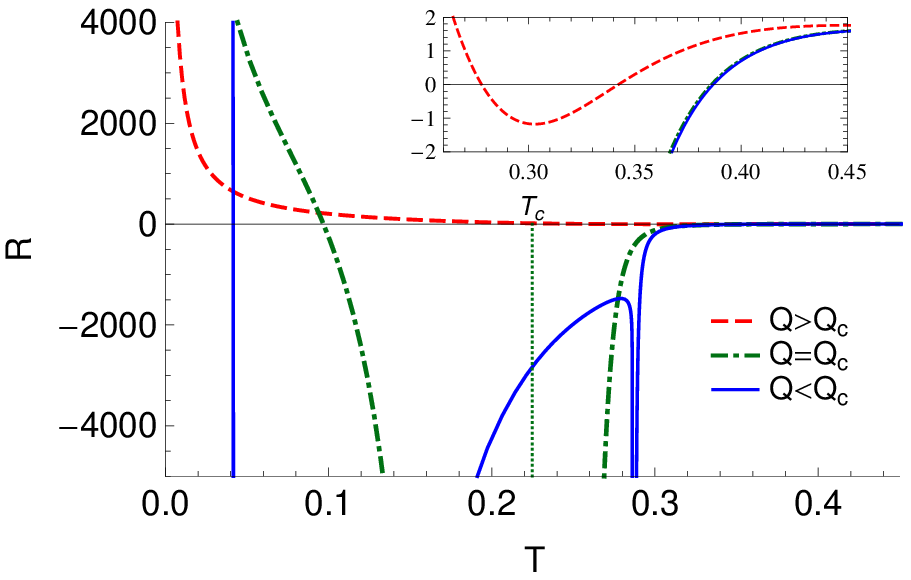}}
\caption{The behavior of Ruppeiner invariant $\mathfrak{R}$ versus $T$ for
linear Maxwell case with $l=b=1$.}
\label{fig3}
\end{figure*}
\begin{figure*}[t]
\centering{%
\subfigure[~$z=2$]{
   \label{fig4a}\includegraphics[width=.46\textwidth]{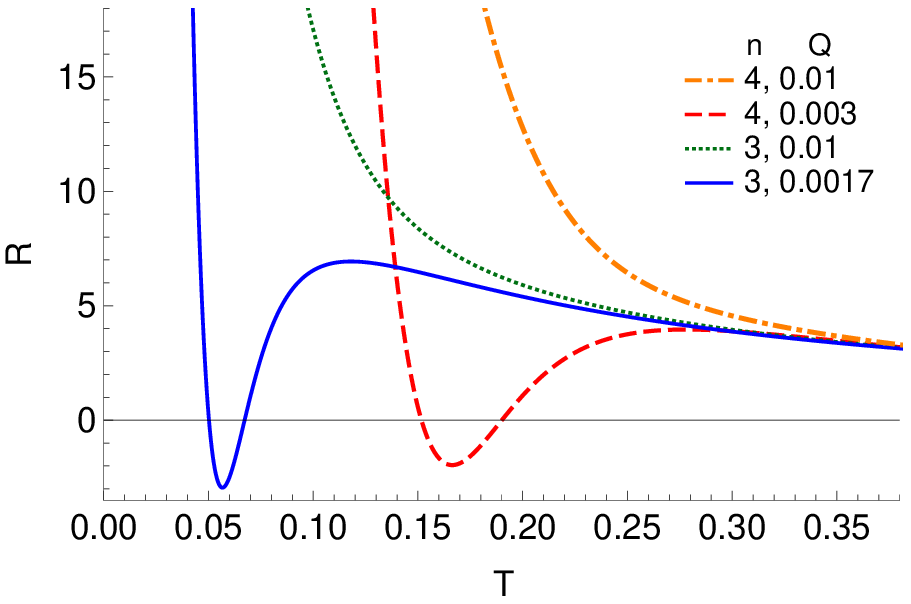}\qquad}} 
\subfigure[~$z=2.5$]{
   \label{fig4b}\includegraphics[width=.46\textwidth]{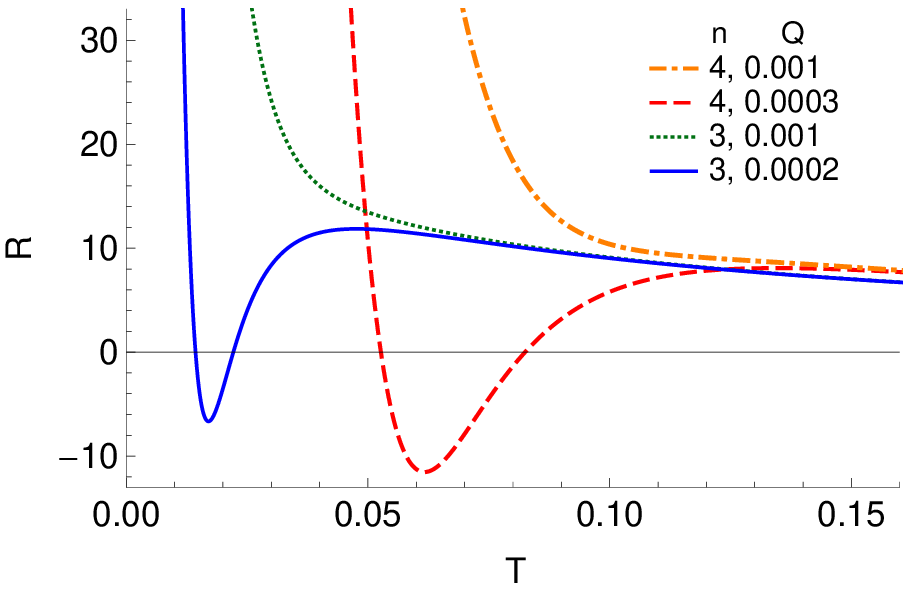}}
\caption{The behavior of Ruppeiner invariant $\mathfrak{R}$ versus $T$ for
linear Maxwell case with $l=b=1$.}
\label{fig4}
\end{figure*}
\begin{figure*}[t]
\centering{%
\subfigure[~$n=3$, $z=1$, $Q=0.049$]{
   \label{fig5a}\includegraphics[width=.46\textwidth]{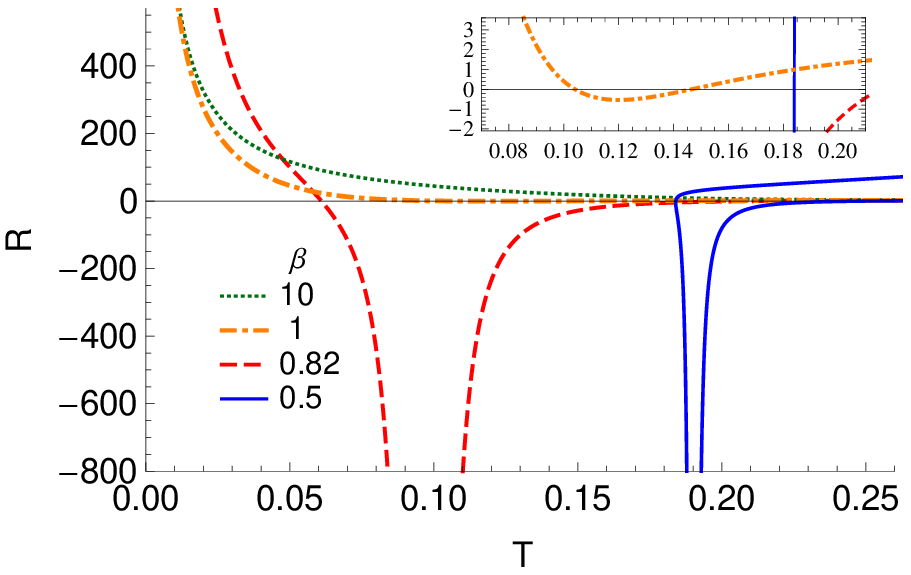}\qquad}} 
\subfigure[~$n=4$, $z=1.5$, $Q=0.02$]{
   \label{fig5b}\includegraphics[width=.46\textwidth]{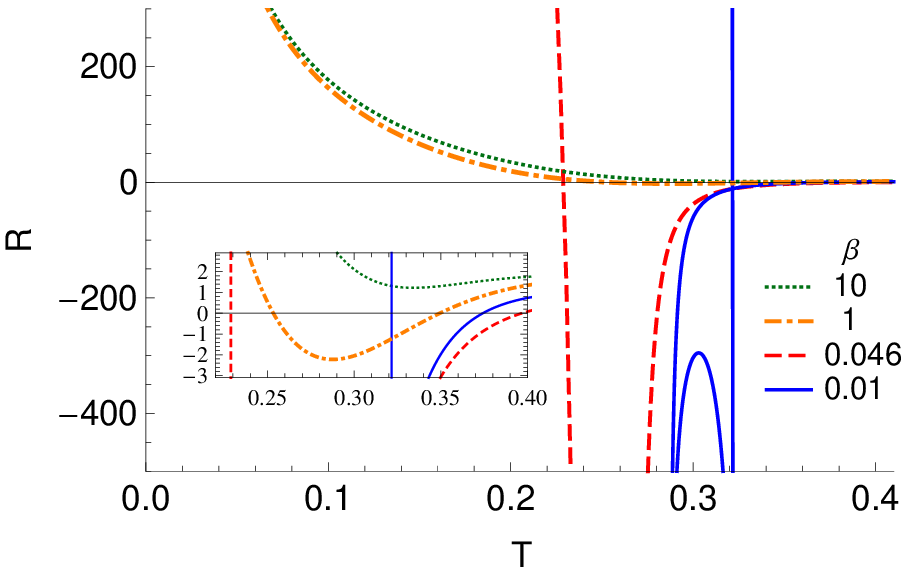}}
\caption{The behavior of Ruppeiner invariant $\mathfrak{R}$ versus $T$ for
Born-Infeld case with $l=b=1$.}
\label{fig5}
\end{figure*}
\begin{figure*}[t]
\centering{%
\subfigure[~$n=3$, $z=1$, $Q=0.049$]{
   \label{fig6a}\includegraphics[width=.46\textwidth]{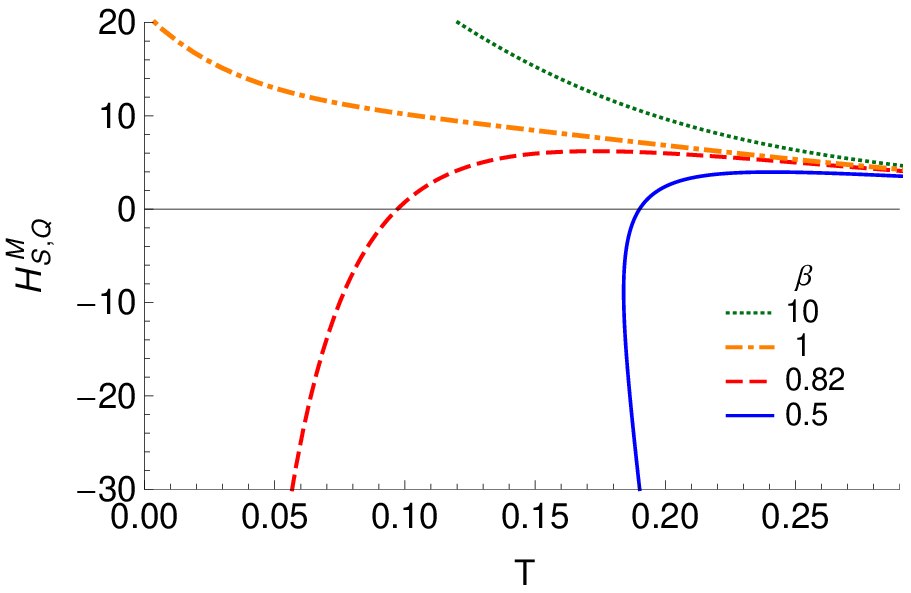}\qquad}} 
\subfigure[~$n=4$, $z=1.5$, $Q=0.02$]{
   \label{fig6b}\includegraphics[width=.46\textwidth]{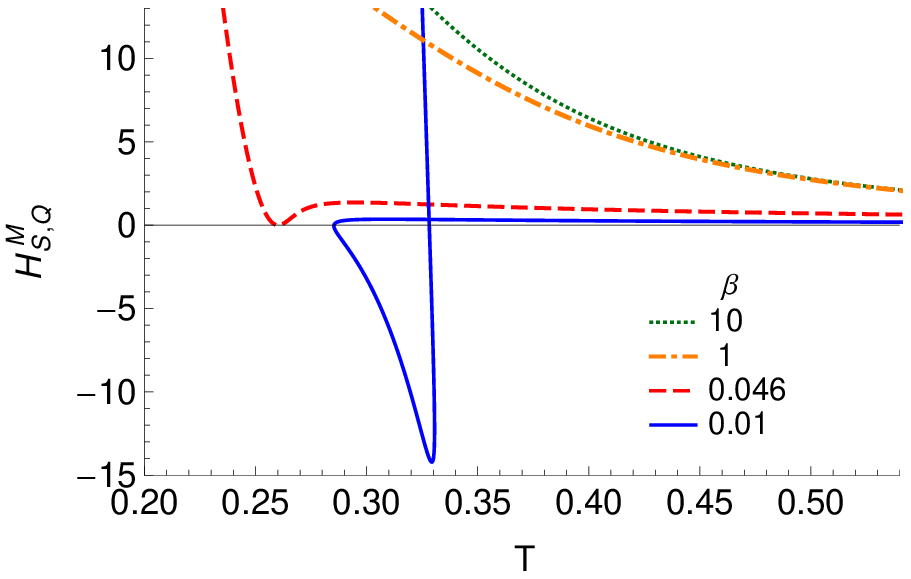}}
\caption{The behavior of determinant of Hessian matrix $\mathbf{H}_{S,Q}^{M}$
versus $T$ for Born-Infeld case with $l=b=1$.}
\label{fig6}
\end{figure*}

We have summarized the above discussion in Figs. \ref{fig3} and \ref{fig4}.
These figures also show the sign of Ruppeiner invariant for different
choices of parameters that determines the type of interaction between black
hole molecules \cite{Rup2,Rup5,Rup6}. Fig. \ref{fig3a} is depicted for
RN-AdS case ($n=3$, $z=1$). In this case, it can be seen that for $Q>Q_{c}$,
the Ruppeiner invariant diverges only for extremal black holes. As Fig. \ref%
{fig3a} shows, there is also a range of $T$ for which $\mathfrak{R}<0$,
namely the dominated interaction between black hole molecules is attractive.
Furthermore, the interactions near zero temperature is the same as
interactions of Fermi gas molecules near zero temperature \cite{Rup2}.
According to Fig. \ref{fig2a}, for $Q>Q_{c}$, $\mathbf{H}_{S,Q}^{M}$ is
positive (also $M_{QQ}>0$ (see (\ref{MQQ}))), and therefore the system is
stable for all $T$ region. For $Q=Q_{c}$, in addition to zero temperature,
we have another temperature ($T_{c}$), that divergence of $\mathfrak{R}$
occurs at it (see Fig. \ref{fig3a}). At zero temperature, the Ruppeiner
invariant goes to $+\infty $ while at $T_{c}$, it goes to $-\infty $. The
latter case is similar to the Van der Waals gas phase transition at critical
point in this sense that in phase transition temperature,\textbf{\ }$%
\mathfrak{R}$ goes to $-\infty $ \cite{Rup1,Rup6}. For $Q=Q_{c}$, $\mathfrak{%
R}$ becomes positive when we get away from second divergence ($T_{c}$) on
temperature axis. In $Q=Q_{c}$ case, $\mathbf{H}_{S,Q}^{M}$ is positive and
just vanishes at $T_{c}$ (Fig. \ref{fig2a}), so, the system is always
thermally stable\textbf{.}\ For $Q<Q_{c}$, there are three divergences; one
at $T=0$, one at $T<T_{c}$ and one at $T>T_{c}$. In this case, according to
Fig. \ref{fig2a},\ $\mathbf{H}_{S,Q}^{M}$ is negative in the temperature
region between two roots and show instability. This not-allowed region is
equivalent to the temperature region between two divergences of Ruppeiner
invariant for\textbf{\ }$Q<Q_{c}$ (Fig. \ref{fig3a}). Figures \ref{fig2b}
and \textbf{\ }\ref{fig3b} show the same properties for black holes with
different parameters. In this case, $T_{c}$ is greater than one of previous
case while $Q_{c}$ is lower. Fig. \ref{fig4} shows the behavior of Ruppeiner
invariant for $z\geq 2$. As this figure shows, there are just divergences at 
$T=0$. The properties of black hole molecular interactions ($\mathfrak{R}>0$%
: Repulsion, $\mathfrak{R}=0$: No interaction and $\mathfrak{R}<0$:
Attraction) depend on parameters such as dimension of space time and charge,
in this case. According to Eq. (\ref{deno}), for $z=2$,\textbf{\ }$\mathbf{H}%
_{S,Q}^{M}$ is always positive. For $z>2$, we can find $Q$ from\textbf{\ }$%
T_{LM}=0$ and put it in\textbf{\ }$\mathbf{H}_{S,Q}^{M}$ to receive 
\begin{equation}
\mathbf{H}_{S,Q,T=0}^{M}=\frac{(n+z-1)b^{2z-2}2^{(n-5)/(1-n)}}{%
(n-1)(n+z-3)l^{2z}S^{(n-2)/(n-1)}}>0.
\end{equation}%
Thus, since $\mathbf{H}_{S,Q}^{M}$ nowhere vanishes for $z>2$ (see
discussions below (\ref{TT})) and is positive at $T=0$ according to above
equation, it is positive throughout the temperature region and therefore
system is always thermally stable for $z>2$.

Regarding the nature of phase transition occurred at zero temperature where
Ruppeiner invariant diverges, we discussed in previous section via
Landau-Lifshitz theory of thermodynamic fluctuations. However, regarding the
phase transitions occurred at divergences of $\mathfrak{R}$ at finite
temperatures, we can give some comments here. We have seen two kinds of
phase transitions here for $z<2$ (see Fig. \ref{fig3}) namely continues (for 
$Q=Q_{c}$ where $\mathfrak{R}$ diverges at just one finite temperature or
entropy) and discontinues (for $Q<Q_{c}$ where $\mathfrak{R}$ diverges at
two finite temperatures or entropies and we have a jump between these two
points since there is no thermally stable black hole between them). Both of
these phase transitions can be considered as small/large black holes phase
transitions. The first reason for this argument is that as temperature
increases, entropy or equivalently size of black hole increases (note that $%
\partial S/\partial T=M_{SS}^{-1}>0$). Therefore, the left side of phase
transition points where temperature is lower, we have small size black holes
and the right side where temperature is higher we have large size ones. This
fact can also be seen from the behavior of Ruppeiner invariant magnitude in
two sides of phase transitions. For small black holes, we expect the finite
correlation between possible black hole molecules (of course far from phase
transition points) because those are close to each other. For large black
holes, we expect the correlation between possible molecules to tend to a
small value near zero since molecules become approximately free. These
expected behaviors can be seen in Fig. \ref{fig3}.

\subsection{Born-Infeld case}

\begin{figure*}[t]
\centering{%
\subfigure[~$n=3$, $z=1$, $Q=0.049$]{
   \label{fig7a}\includegraphics[width=.46\textwidth]{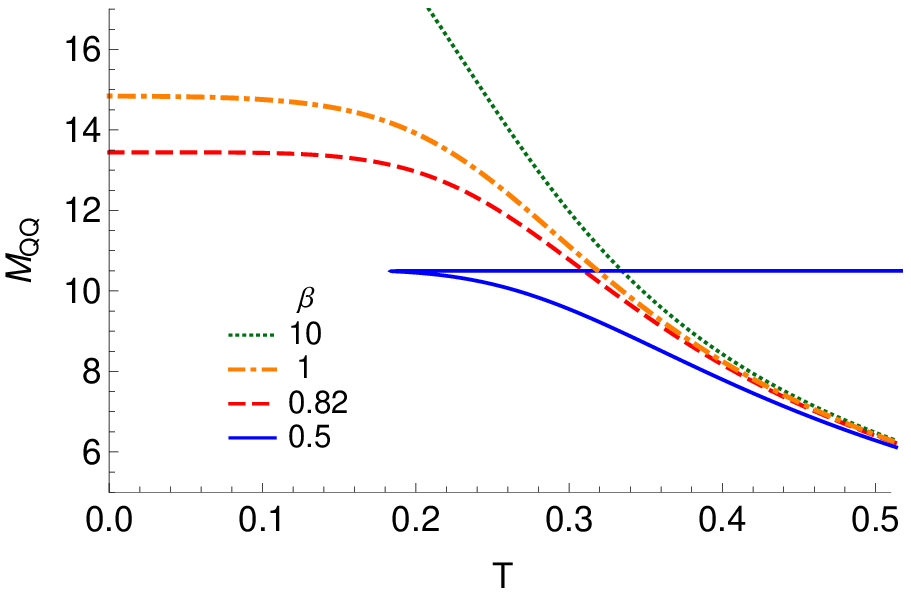}\qquad}} 
\subfigure[~$n=4$, $z=1.5$, $Q=0.02$]{
   \label{fig7b}\includegraphics[width=.46\textwidth]{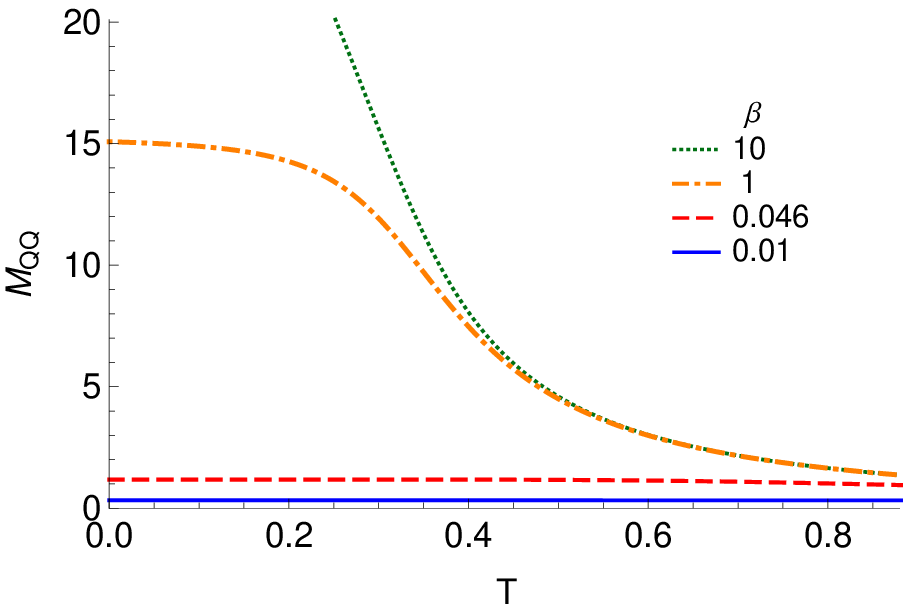}}
\caption{The behavior of $M_{QQ}$ versus $T$ for Born-Infeld case with $%
l=b=1 $.}
\label{fig7}
\end{figure*}

\begin{figure*}[t]
\centering{%
\subfigure[]{
   \label{fig8a}\includegraphics[width=.46\textwidth]{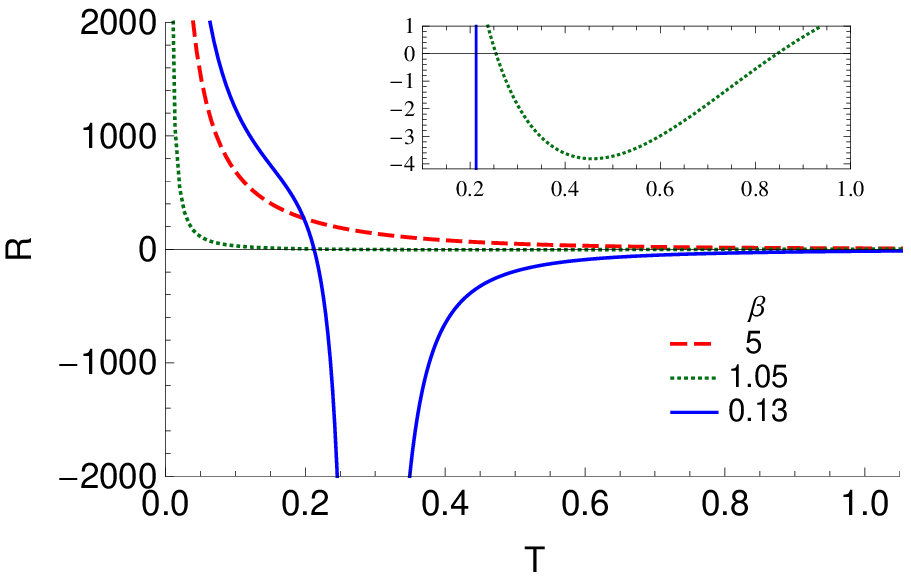}}} \newline
\centering{\ 
\subfigure[]{
   \label{fig8b}\includegraphics[width=.46\textwidth]{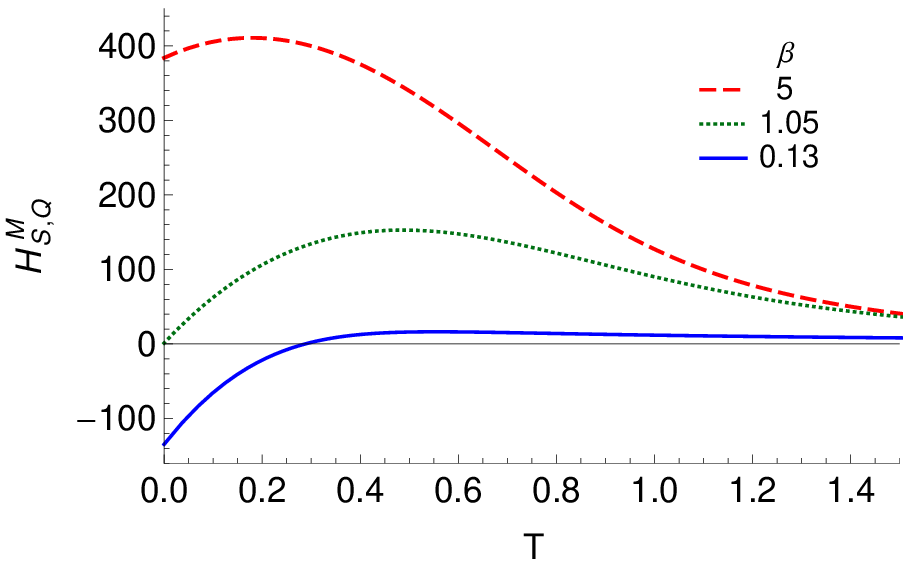}\qquad} 
\subfigure[~]{
    \label{fig8c}\includegraphics[width=.46\textwidth]{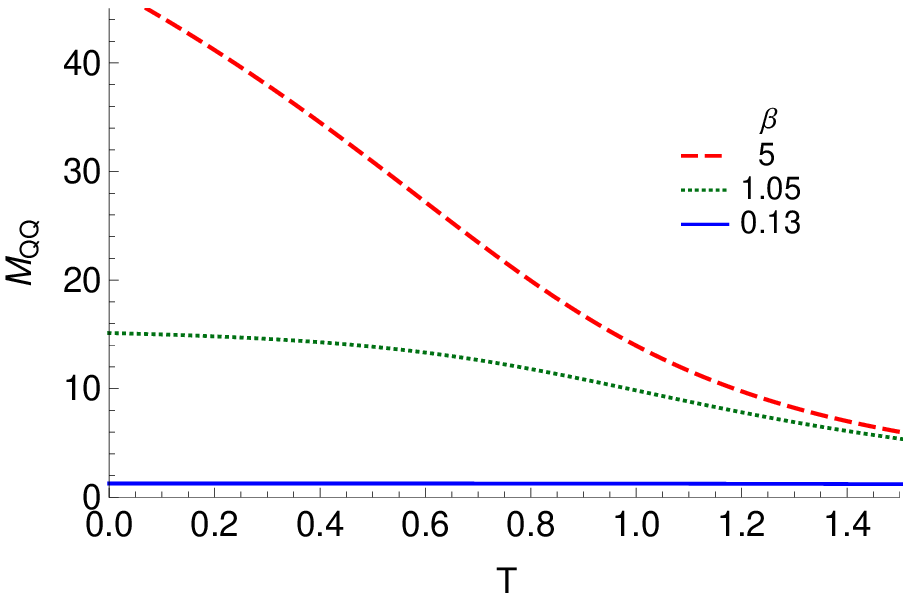}}}
\caption{The behavior of Ruppeiner invariant $\mathfrak{R}$, $\mathbf{H}%
_{S,Q}^{M}$ and $M_{QQ}$ versus $T$ for Born-Infeld case with $b=0.9$, $%
l=0.76$, $n=6$, $z=3$ and $Q=0.018$.}
\label{fig8}
\end{figure*}

\begin{figure*}[t]
\centering\includegraphics[width=.46\textwidth]{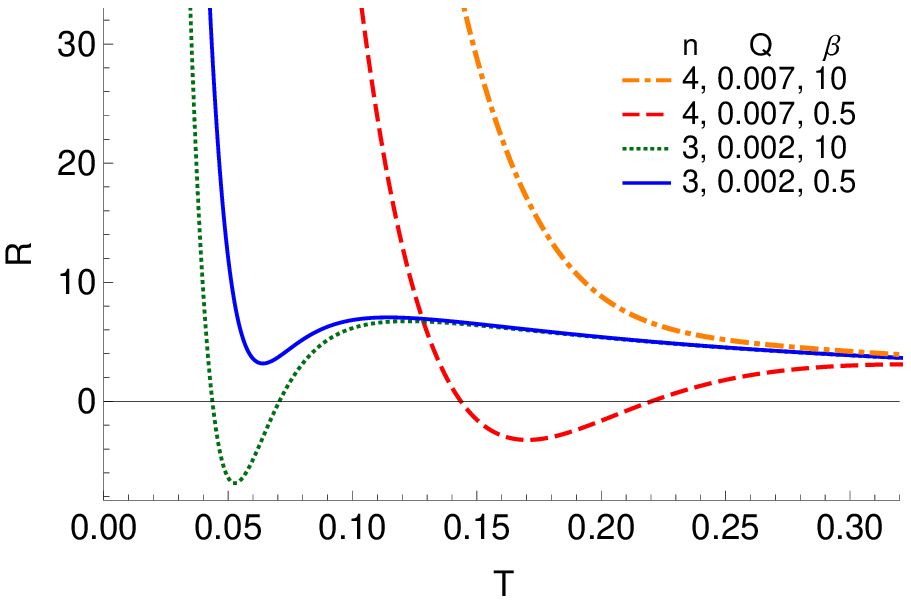}
\caption{The behavior of Ruppeiner invariant $\mathfrak{R}$ versus $T$ for
Born-Infeld case with $z=2$ and $l=b=1$.}
\label{fig9}
\end{figure*}

\begin{figure*}[t]
\centering{%
\subfigure[]{
   \label{fig10a}\includegraphics[width=.46\textwidth]{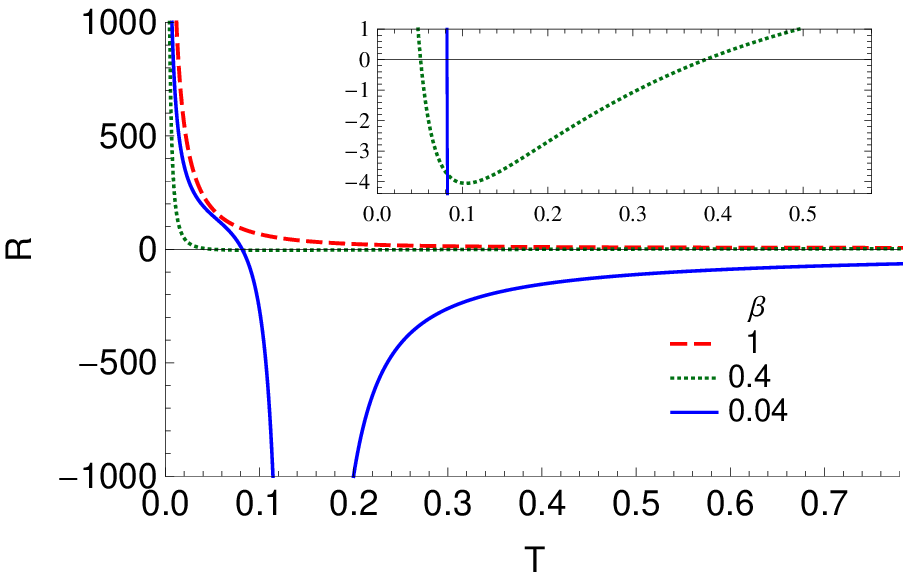}\qquad}} 
\subfigure[]{
   \label{fig10b}\includegraphics[width=.46\textwidth]{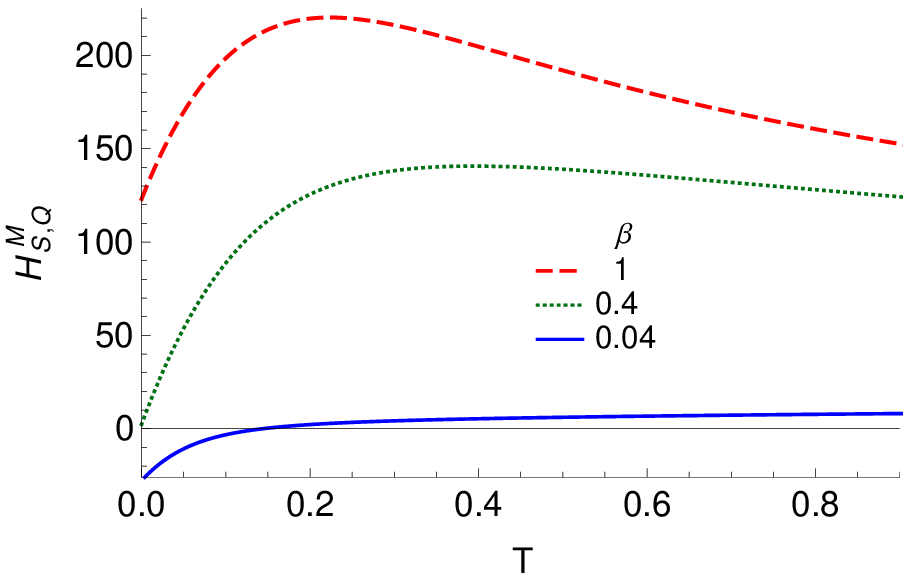}}
\caption{The behavior of Ruppeiner invariant $\mathfrak{R}$ and $\mathbf{H}%
_{S,Q}^{M}$ versus $T$ for Born-Infeld case with $b=0.91$, $l=0.72$, $n=3$, $%
z=4$ and $Q=0.012$.}
\label{fig10}
\end{figure*}

For Born-Infeld case, we can calculate the Ruppeiner invariant by using Eqs.
(\ref{Temp}), (\ref{M}) and (\ref{MetGTD}). The Ruppeiner invariant in this
case is very complicated due to the presence of hypergeometric functions.
Therefore, in this case we discuss the thermodynamic geometry
non-analytically and by looking at plots. We study $z<2$, $z>2$, $z=2$ and $%
z=n+1$ cases separately. First, we study $z<2$ case. Fig. \ref{fig5a} shows
that changing $\beta $ can cause change in dominated interaction. For
instance, in a range of $T$, we have negative $\mathfrak{R}$ (attraction)
for $\beta =1$ (note that in this range system is thermally stable as one
can see from Figs. \ref{fig6a} and \ref{fig7a}). For $\beta =1$, the system
behaves like Fermi gas in zero temperature namely $\mathfrak{R}$ goes to
positive infinity at zero temperature \cite{Rup2}. For $\beta =0.82$,
Ruppeiner invariant diverges at two points that one of them is zero
temperature. According to Fig. \ref{fig6a}, for temperatures lower than the
second divergence point, $\mathbf{H}_{S,Q}^{M}$ is negative and therefore
system is thermally unstable. Since $M_{QQ}>0$ (Fig. \ref{fig7a}), the
system is thermally stable just for temperatures greater than the
temperature of second divergence for $\beta =0.82$. Fig. \ref{fig5a} shows
that there is no extremal black hole for $\beta =0.5$ i.e. we have a black
hole with just single horizon. The allowed temperature region is
temperatures greater than the temperature of divergence according to Figs. %
\ref{fig6a} and \ref{fig7a}. In Figs. \ref{fig5b}, \ref{fig6b} and \ref%
{fig7b}, respectively Ruppeiner invariant, $\mathbf{H}_{S,Q}^{M}$ and $%
M_{QQ} $ are depicted for different choices of parameters. It is remarkable
to mention that, in the case of $\beta =0.046$, Fig. \ref{fig5b} shows that
the behavior of system looks like Van der Waals gas at phase transition
temperature i.e. $\mathfrak{R}$ goes to negative infinity at this point \cite%
{Rup1,Rup6}. For $z>2$, the behavior of $\mathfrak{R}$ is depicted in Fig. %
\ref{fig8a}. It can be seen that the type of dominated interaction changes
for different $\beta $'s and we have negative $\mathfrak{R}$ for some cases.
In this case, we have a behavior like Fermi gas at zero temperature for
extremal black holes. For $\beta =0.13$, there is a divergence at non-zero
temperature that for temperatures lower than it, system is unstable (Figs. %
\ref{fig8b} and \ref{fig8c}). In the case of\textbf{\ }$z=2$, $\mathbf{H}%
_{S,Q}^{M}$ and $M_{QQ}$ are 
\begin{equation}
\left. \mathbf{H}_{S,Q}^{M}\right\vert _{z=2}=\frac{(n+1)b^{2}}{%
2^{(n-5)/(n-1)}l^{4}(n-1)^{2}S^{2n/(n-1)}\Gamma },
\end{equation}%
and%
\begin{equation*}
\left. M_{QQ}\right\vert _{z=2}=\frac{b^{2}\pi }{(n-1)lS\Gamma },
\end{equation*}%
which are always positive and therefore system is always stable and $%
\mathfrak{R}$ experiences no divergence (Fig. \ref{fig9}). In this case, for
different values of nonlinear parameter $\beta $, we have different
dominated interaction. For this case, possible molecules of black hole
behave like Fermi gas at zero temperature. The last case is $z=n+1$. In this
case $M_{QQ}$ is 
\begin{equation}
\left. M_{QQ}\right\vert _{z=n+1}=\frac{b^{2n}\beta ^{2}\left( \Gamma
-1\right) }{4\pi (n-1)Q^{2}l^{n}\Gamma },
\end{equation}%
which is positive for all temperatures. The behavior of Ruppeiner invariant
and $\mathbf{H}_{S,Q}^{M}$ are depicted in Figs. \ref{fig10a} and \ref%
{fig10b} for this case, respectively. As one can see the type of interaction
is $\beta $-dependent for some temperatures. For $\beta =0.04$, $\mathbf{H}%
_{S,Q}^{M}$ is positive just for temperatures greater than the finite
temperature of divergence (Fig. \ref{fig10b}) and therefore system is
thermally stable for this range of temperatures.

Most of phase transitions discovered above in the presence of BI
electrodynamics at finite temperatures, cannot be interpreted as small/large
black hole phase transitions because in these cases small size black holes
are unstable. Further studies to disclose the nature of these phase
transitions are called for.

\section{Summary and closing remarks\label{Sum}}

In many condensed matter systems, fixed points governing the phase
transitions respect dynamical scaling $t\rightarrow \lambda ^{z}t$, $\vec{%
\mathbf{x}}\rightarrow \lambda \vec{\mathbf{x}}$ where $z$ is dynamical
critical exponent. The gravity duals of such systems are Lifshitz black
holes. In this paper, we first sought for the ($n+1$)-dimensional
Born-Infeld (BI) charged Lifshitz black hole solutions in the context of
dilaton gravity. We found out that these solutions are different for the
cases $z=n+1$ and $z\neq n+1$. We obtained both these solutions and showed
that the solution for the case $z=n+1$ can never be Schwartzshild-like.
Then, we studied thermodynamics of both cases by calculating conserved and
thermodynamical quantities and checking the satisfaction of first law of
thermodynamics. After that, we looked for the Hawking-Page phase transition
for our solutions, both in the cases of linearly and BI charged black holes.
We studied this phenomenon and effects of different parameters on it by
presenting the behaviors of temperature $T$ with respect to entropy $S$ at
fixed electrical potential energy $U$ and also Gibbs free energy $G$ with
respect to $T$. Then, we turned to discuss the phase transitions inside the
black holes. In this part, we first presented the improved Davies quantities
that show the phase transition points coincided with ones of Ruppeiner
geometry. This coincidence has been proved directly in appendix \ref{app1}.
All of our solutions provided that those are thermally stable at zero
temperature show the divergence at this point both from Ruppeiner and Davies
points of view. Using Landau-Lifshitz theory of thermodynamic fluctuations,
we showed that this phase transition is a transition on radiance properties
of black holes. At zero temperature, an extreme black hole can just radiate
through superradiant scattering whereas a nonextreme black hole at finite
temperature can give off particles and radiation via both spontaneous
Hawking radiation and superradiant scattering.

Next, we turned to study Ruppeiner geometry for our solutions. We
investigated thermal stability, interaction type of possible black hole
molecules and phase transitions of our solutions for linearly and
nonlinearly BI charged cases separately. For linearly charged case, we
showed that there are no diverging points for Ricci scalar of Ruppeiner
geometry (Ruppeiner invariant) at finite temperature for the case $z\geq 2$.
For $z<2$, it was found that the number of divergences (which show the phase
transitions) at finite temperatures depend on the value of charge $Q$. We
introduced a critical value for charge $Q_{c}$ that for values greater than
it there is no divergence at finite temperature, for values lower than it
there are at most two divergences and for $Q=Q_{c}$, there is just one
diverging point for Ruppeiner invariant. For the case of $Q<Q_{c}$, there is
a thermally unstable region for systems between two divergences at finite
temperatures. So, this phase transition can be claimed as a discontinues
phase transition between small and large black holes. For small black holes
not close to transition point, we observed finite magnitude for Ruppeiner
invariant $\mathfrak{R}$. This is reasonable since the magnitude of $%
\mathfrak{R}$ shows the correlation of possible black hole molecules. Also,
for large black holes the magnitude of Ruppeiner invariant tends to a very
small value as expected. For $Q=Q_{c}$, the solutions show a continues
small/large black holes phase transition at finite temperature. In the case
of BI charged solutions, we investigated the Ruppeiner geometry and thermal
stability for $z<2$, $z>2$, $z=2$ and $z=n+1$ separately. In some of these
cases, small black holes were thermally unstable. So, more studies are
called for to discover the nature of phase transitions at diverging points
of $\mathfrak{R}$. In both linearly and nonlinearly charged cases, for some
choices of parameters, the black hole system behaves like a Van der Waals
gas near transition point.

Finally, we would like to suggest some related interesting issues which can
be considered for future studies. It is interesting to repeat the studies
here such as Hawking-Page phase transition, Ruppeiner geometry and
Landau-Lifshitz theory for black branes to discover the effect of different
constant curvatures of ($n-1$)-dimensional hypersurface on those phenomena.
One can also seek for any signature of different phase transitions
discovered here such as Hawking-Page phase transition, and phase transitions
determined by Ruppeiner geometry, in dynamical properties of solutions by
investigating quasi-normal modes. Some of these works are in progress by
authors.

\begin{acknowledgments}
We are grateful to Prof. M. Khorrami for very useful discussions. MKZ would
like to thank Shanghai Jiao Tong University for the warm hospitality during
his visit. Also, MKZ, AD and AS thank the research council of Shiraz
University. This work has been financially supported by the Research
Institute for Astronomy \& Astrophysics of Maragha (RIAAM), Iran.
\end{acknowledgments}

\appendix

\section{Suitable thermodynamic quantities to determine phase transitions 
\label{app1}}

In \cite{RupGeo1} and \cite{RupGeo5}, authors have shown that the
divergences of specific heat at constant electrical potential, $C_{U}$,
analog of volume expansion coefficient, $\alpha $, and analog of isothermal
compressibility coefficient $\kappa _{T}$ are in coincident with the phase
transitions specified by Ruppeiner invariant. The definition of these
quantities are%
\begin{equation}
C_{U}=T\left( \frac{\partial S}{\partial T}\right) _{U}\text{, \ }\alpha =%
\frac{1}{Q}\left( \frac{\partial Q}{\partial T}\right) _{U}\text{ \ and }%
\kappa _{T}=\frac{1}{Q}\left( \frac{\partial Q}{\partial U}\right) _{T}\text{%
.}
\end{equation}%
Here we will prove that these quantities are exactly suitable ones to
characterize phase transitions shown by Ruppeiner invariant. We showed in
section \ref{RG} that divergences of Ruppeiner invariant occurs at roots of
determinant of Hessian matrix $\mathbf{H}_{S,Q}^{M}=M_{SS}M_{QQ}-M_{SQ}^{2}$
and also zero temperature. In our proof, we will show that $\mathbf{H}%
_{S,Q}^{M}$ exactly exist at denominator of all above suitable thermodynamic
quantities.

Let us start with $C_{U}$. We have%
\begin{equation}
\left. \frac{\partial T\left( S,Q\left( U,S\right) \right) }{\partial S}%
\right\vert _{U}=\left. \frac{\partial T}{\partial S}\right\vert _{Q}+\left. 
\frac{\partial T}{\partial Q}\right\vert _{S}\left. \frac{\partial Q}{%
\partial S}\right\vert _{U}.
\end{equation}%
On the other hand we know that

\begin{equation}
\left. \frac{\partial Q}{\partial S}\right\vert _{U}=-\left. \frac{\partial Q%
}{\partial U}\right\vert _{S}\left. \frac{\partial U}{\partial S}\right\vert
_{Q}.
\end{equation}%
With above relations in hand, one can show that

\begin{eqnarray}
\left. \frac{\partial T\left( S,Q\left( U,S\right) \right) }{\partial S}%
\right\vert _{U} &=&\left. \frac{\partial T}{\partial S}\right\vert
_{Q}-\left. \frac{\partial T}{\partial Q}\right\vert _{S}\left. \frac{%
\partial Q}{\partial U}\right\vert _{S}\left. \frac{\partial U}{\partial S}%
\right\vert _{Q}=\left. \frac{\partial T}{\partial S}\right\vert _{Q}-\frac{%
\left. \frac{\partial T}{\partial Q}\right\vert _{S}\left. \frac{\partial U}{%
\partial S}\right\vert _{Q}}{\left. \frac{\partial U}{\partial Q}\right\vert
_{S}}  \notag \\
&=&\frac{\left. \frac{\partial T}{\partial S}\right\vert _{Q}\left. \frac{%
\partial U}{\partial Q}\right\vert _{S}-\left. \frac{\partial T}{\partial Q}%
\right\vert _{S}\left. \frac{\partial U}{\partial S}\right\vert _{Q}}{\left. 
\frac{\partial U}{\partial Q}\right\vert _{S}}=\frac{M_{QQ}M_{SS}-M_{SQ}^{2}%
}{M_{SS}}=\frac{\mathbf{H}_{S,Q}^{M}}{M_{SS}}.  \label{A1}
\end{eqnarray}%
In the last line of (\ref{A1}), we have used (\ref{intqua}). Eq. (\ref{A1})
shows that $\mathbf{H}_{S,Q}^{M}=M_{SS}M_{QQ}-M_{SQ}^{2}$ is in denominator
of $C_{U}=T\left( \partial S/\partial T\right) _{U}$ and therefore it
exactly diverges at the point where Ruppeiner invariant diverges. To show
this fact for $\alpha $, we should obtain

\begin{equation}
\left. \frac{\partial T\left( Q,S\left( U,Q\right) \right) }{\partial Q}%
\right\vert _{U}=\left. \frac{\partial T}{\partial Q}\right\vert _{S}+\left. 
\frac{\partial T}{\partial S}\right\vert _{Q}\left. \frac{\partial S}{%
\partial Q}\right\vert _{U}.
\end{equation}%
As we know

\begin{equation}
\left. \frac{\partial S}{\partial Q}\right\vert _{U}=-\left. \frac{\partial S%
}{\partial U}\right\vert _{Q}\left. \frac{\partial U}{\partial Q}\right\vert
_{S},
\end{equation}%
and therefore we have

\begin{eqnarray}  \label{A2}
\left. \frac{\partial T\left( Q,S\left( U,Q\right) \right) }{\partial Q}%
\right\vert _{U} &=&\left. \frac{\partial T}{\partial Q}\right\vert
_{S}-\left. \frac{\partial T}{\partial S}\right\vert _{Q}\left. \frac{%
\partial S}{\partial U}\right\vert _{Q}\left. \frac{\partial U}{\partial Q}%
\right\vert _{S}=\left. \frac{\partial T}{\partial Q}\right\vert _{S}-\frac{%
\left. \frac{\partial T}{\partial S}\right\vert _{Q}\left. \frac{\partial U}{%
\partial Q}\right\vert _{S}}{\left. \frac{\partial U}{\partial S}\right\vert
_{Q}}  \notag \\
&=&\frac{\left. \frac{\partial T}{\partial Q}\right\vert _{S}\left. \frac{%
\partial U}{\partial S}\right\vert _{Q}-\left. \frac{\partial T}{\partial S}%
\right\vert _{Q}\left. \frac{\partial U}{\partial Q}\right\vert _{S}}{\left. 
\frac{\partial U}{\partial S}\right\vert _{Q}}=-\frac{M_{QQ}M_{SS}-M_{SQ}^{2}%
}{M_{SQ}}=-\frac{\mathbf{H}_{S,Q}^{M}}{M_{SQ}}.  \notag \\
\end{eqnarray}%
The above relation shows that $\alpha =Q^{-1}\left( \partial Q/\partial
T\right) _{U}$ diverges at the point where Ruppeiner invariant does. At
final, to receive similar result for $\kappa _{T}$, we obtain

\begin{equation}
\left. \frac{\partial U}{\partial Q}\right\vert _{T}=-\left. \frac{\partial U%
}{\partial T}\right\vert _{Q}\left. \frac{\partial T}{\partial Q}\right\vert
_{U}=-\frac{1}{Q\alpha }\left. \frac{\partial U}{\partial T}\right\vert _{Q}.
\end{equation}%
Above relation shows that $\kappa _{T}=Q^{-1}\left( \partial Q/\partial
U\right) _{T}$ is proportional to $\alpha $ and therefore diverges at the
same points as Ruppeiner invariant.

\end{document}